\documentclass[journal]{IEEEtran}
\normalsize
\ifCLASSINFOpdf
\else

\fi

\usepackage{balance}
\usepackage{mathtools}
\usepackage{graphicx}
\usepackage{epstopdf}
\usepackage{amsmath}
\usepackage{amssymb}
\usepackage{amsmath}
\usepackage{cite}
\usepackage{multirow}
\usepackage{booktabs}
\usepackage{subfigure}
\usepackage{array}
\usepackage{lipsum}                     
\usepackage{xargs}                      
\usepackage[pdftex,dvipsnames]{xcolor}  
\usepackage[colorinlistoftodos,prependcaption,textsize=tiny]{todonotes}
\usepackage{siunitx}

\date{}
\usepackage{mdwmath}
\usepackage{blindtext}
\usepackage{eqparbox}
\usepackage{color}

\begin{document}
\title{Analysis of Indoor Uplink Optical Communication Positioning System Exploiting Multipath Reflections} 
\author{\IEEEauthorblockN{Hamid Hosseinianfar\IEEEauthorrefmark{1} and Maite Brandt-Pearce\IEEEauthorrefmark{3}\\}
	\IEEEauthorblockA{Charles L. Brown Department of Electrical and Computer Engineering, \\ University of Virginia, Charlottesville, VA 22904.\\}
		Email: \IEEEauthorrefmark{1}hh9af@virginia.edu,
		\IEEEauthorrefmark{3}mb-p@virginia.edu}
\maketitle
\begin{abstract}
In this paper, we introduce an uplink optical wireless positioning system for indoor applications. This technique uses fingerprints based on the indoor optical wireless channel impulse response for localization. Exploiting the line of sight peak power (LOS), the second power peak (SPP) of the impulse response, and the delay between the LOS and SPP, we present a proof of concept design and theoretical analysis for localization employing a single fixed reference point, i.e., a photodetector (PD) on the ceiling. Adding more PDs leads to more accurate transmitter position estimation. As a benchmark, we present analytical expressions of the Cramer-Rao lower bound (CRLB) for different numbers of PDs and features. We further present closed form analytical approximations for the chosen features of the channel impulse response. Simulation results show a root mean square (RMS) positioning accuracy of 25\,cm and 5\,cm for one and four PDs, respectively, for a typical indoor room at high SNR. Numerical results verify that the derived analytic approximations closely match the simulations.

\end{abstract}


\IEEEpeerreviewmaketitle

\section{Introduction}
  
There is a high demand nowadays for localization services in many applications like robotics, unmanned aerial vehicles, internet of things applications and self-driving cars. While GPS provides a robust localization service for outdoor applications, its poor coverage for indoor environments, as well as its poor accuracy indoors, leads to a strong need for a robust indoor localization system. In this paper, we address this problem by introducing a new optical wireless technique that uses the features of an infrared (IR) uplink channel impulse response for positioning. 

The exponential growth of light-emitting diode (LED) illumination infrastructures alongside the high network throughput capacity possible with visible light communication systems (VLC) has led to a growing interest in VLC as the next generation of network access points \cite{noshad2013can}. On the other hand, optical wireless techniques can also provide centimeter accuracy for indoor localization services and have been considered as one of the most promising indoor localization approaches \cite{zhang2013comparison, 6685759,wang2013position}. Using the same VLC infrastructure paves the way for introducing visible light positioning systems for either localization purposes or simply to assist communication services, such as handover and resource allocation\cite{rahaim2012state}. Previous research has shown that enhanced handover decisions based on the knowledge of user locations and motion tracking can improve the overall quality of service (QoS), when compared to techniques that monitor signal strength alone, by reducing the number of unnecessary channel transfers\cite{hou2006vertical,pollini1996trends}. 

Visible light positioning methods can generally be classified into three groups: proximity, triangulation, and fingerprinting. Proximity techniques give an estimation of the approximate location of users, i.e., the closest access point \cite{do2016depth}. Triangulation methods rely on one of three features of the received signals: time of arrival (TOA), angle of arrival (AOA), or received signal strength (RSS). 
RSS-based techniques, which use the intensity of the signal for localization, can achieve a high accuracy in visible light positioning systems due to the strong line of sight (LOS) signal in wireless optical systems. However, the accuracy of these techniques is limited due to its poor performance in shadowing and multi-path environments, which make the relationship between the distance and RSS unpredictable \cite{zhang2014theoretical, zhang2013comparison,csahin2015hybrid}.
TOA-based techniques rely on the arrival time of the signal from different transmitters for estimation and, hence, require perfect synchronization between the transmitters, which can add complexity and limit the application of these systems. Theoretical limits have been presented on the accuracy of TOA-based \cite{wang2013position} and RSS-based \cite{6880333} positioning techniques. AOA localization techniques use the angle of arrival of the LOS signal from different transmitters for localization. 
They can estimate the user's location using one imaging receiver when the height of the user is known and two imaging receivers when it is not \cite{zhang2013comparison}.
AOA has been used to locate users with an accuracy of $5$ cm in practice \cite{5381102}, by measuring the angle at which the line-of-sight (LOS) signal from the transmitter is received. Fingerprinting methods estimate the relative location of the user by matching real-time measurements with a previously-collected fingerprint map. Pattern recognition techniques such as probabilistic methods, k-nearest-neighbors (k-NN), and correlation have been studied for fingerprinting methods based on the downlink signal \cite{lin2005performance,7795368}. However, the computational complexity is a challenging aspect of fingerprinting algorithms, especially for implementation on portable devices.

In this paper, an uplink fingerprint-based indoor optical localization algorithm is proposed. While other VLC techniques consider the multipath signal as noise, our localization technique, which was first introduced in \cite{7996815}, uses the characteristics of the optical channel impulse response to locate users. In this paper, we expand on \cite{7996815} by presenting an expression of the Cramer Rao lower bound (CRLB) for the proposed fingerprinting approach. We also examine the performance of the proposed techniques for a finite transmitter bandwidth (BW). This technique can estimate the user location using a single reference point, i.e., one PD, which not only reduces the complexity of the system by employing a lower number of sensors on the ceiling, but also makes the localization possible in severe shadowing. To the authors' knowledge, ours is the only technique in the published literature that can estimate the user location using a single measurement. However, adding more reference points in our algorithm enhances the positioning accuracy. The algorithm performance is first evaluated for infinite system BW to show the best possible accuracy for a perfect channel impulse response. We then look at the usefulness of impulse response features when taking the finite LED BW into account.

In contrast to most research on VLC indoor positioning that applies a user-side localization technique \cite{wang2013position}, the proposed algorithm is developed for the infrared (IR) uplink of a VLC system. Using uplink fingerprinting localization not only reduces the processing load and hardware complexity on the user-side, but also provides a more secure approach, in the sense that the physical infrastructure information need not be shared with users. Although network-side localization can be seen as violating users' privacy by monitoring them, the security-privacy trade-off is advantageous for specific applications such as robot-navigation, network multiple access (MAC) layer optimization, and fire-hazard monitoring, in which system monitoring of the number and the location of users is required.

The rest of the paper is organized as follows. In Section \ref{System_Description}, the system model and measurement scenario are discussed. The positioning algorithm is presented in Section \ref{PA}. CRLB calculations are derived in Section \ref{Sec:CRLB}. Numerical results are presented and discussed in Section \ref{Sec:Mum}. Finally, the paper is concluded in Section \ref{Concl}.
\section{System Description}
\label{System_Description}
In this section, we first describe an overview of the proposed system. We then discuss the channel model, fingerprinting, and coding aspects of our system design.
\begin{figure}[!t]
 \centering
 \includegraphics[width=3.1in]{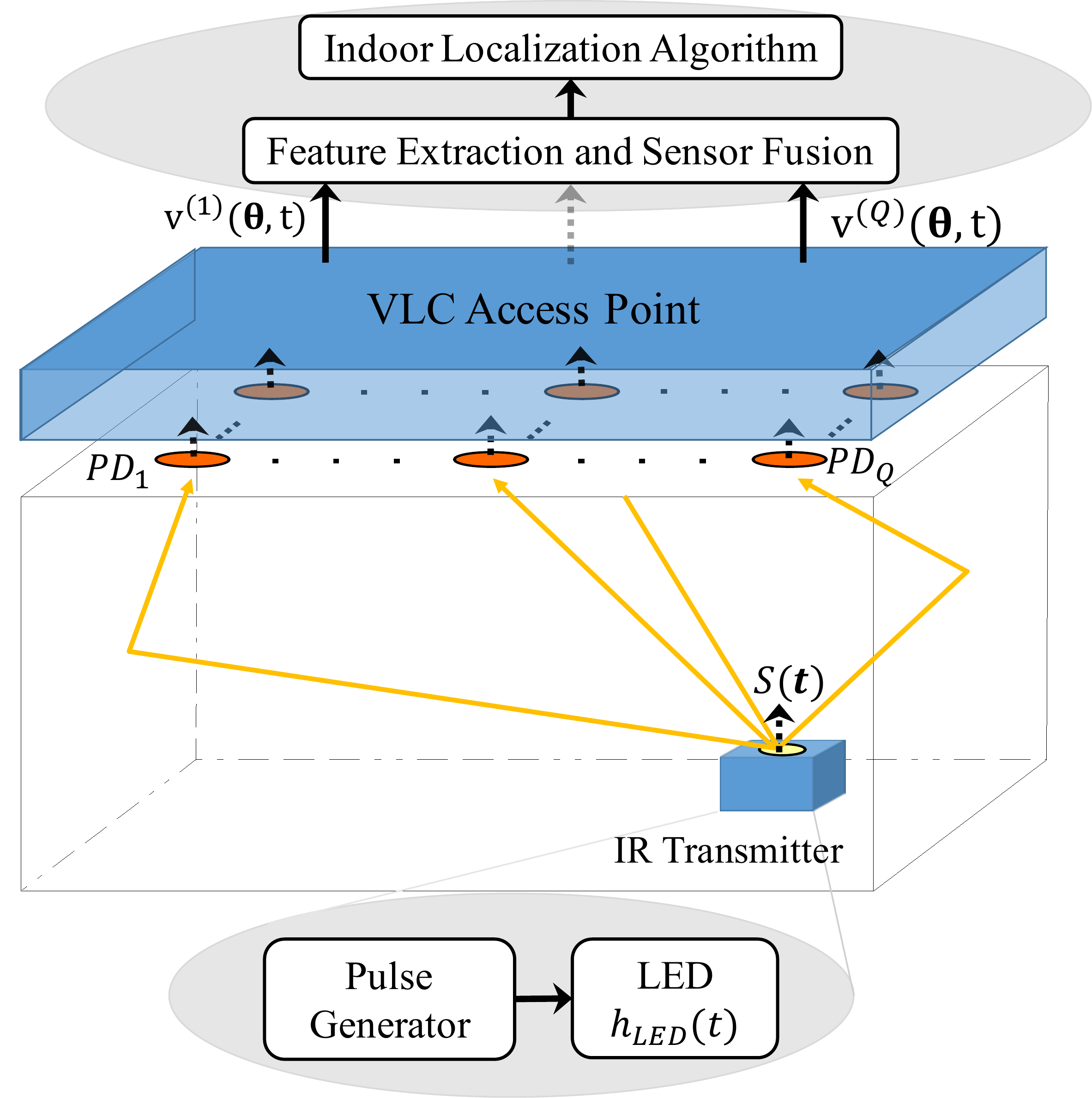} 
 \caption{System configuration for visible light communication uplink system and the impulse responses based on it. }
 \label{fig:room}
\end{figure}
\subsection{System Overview}
The proposed indoor localization system is considered part of a visible light communication system in which there are white LED fixtures on the ceiling that transmit downlink data and infrared photo-detectors (PDs) that capture uplink signals, as shown in Fig.~\ref{fig:room}. In this model, the $q$th PD is assumed to be installed on the ceiling at position $(x^{(q)}, y^{(q)}, z^{(q)}), q \in\left \{  1, \cdots, Q\right \},$ facing vertically downwards, and the user located at coordinates $(x, y, z)$ is assumed to be at a known height $z$ and have an infrared LED transmitter that is facing vertically upwards. 

The key idea in this work is to use the channel impulse responses of the uplink channels to locate the user by mapping characteristics to the location of the user. Our work assumes a single user in the room (with generalization to multiple users discussed below). The transmitter is required to transmit a series of narrow time pulses. On the network-side, the receivers capture the channel impulse responses and, for each pulse, extract the significant features of the received signal.    

\subsection{Channel Model} 
\begin{figure}[!t]
 \centering
 \includegraphics[width=3.3in]{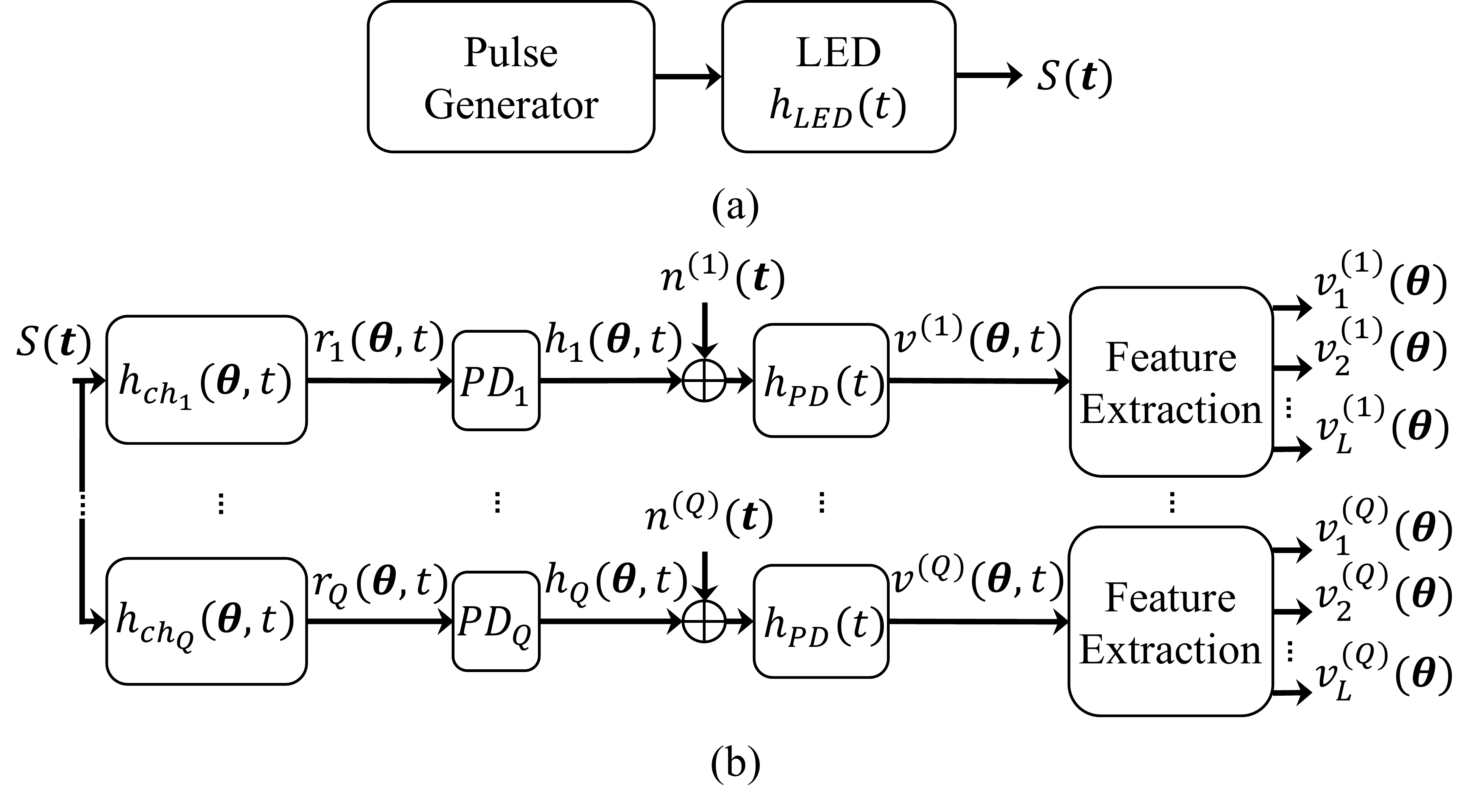} 
 \caption{Block diagram of system. a) transmitter and b) receiver structures, including the optical wireless channel between transmitter and each of receivers.}
 \label{fig:System_BD}
\end{figure}
\begin{figure}[!t]
 \centering
 \includegraphics[width=3.2in]{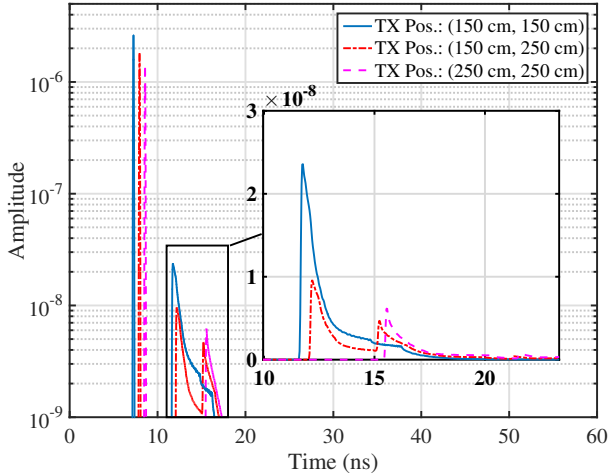} 
 \caption{Impulse response of the channel for different locations of the transmitter when the receiver is located at $(150 ~\si{cm}, \ 150~\si{cm}, 300~\si{cm})$ and the simulation parameters are as in Table \ref{TableI}.}
 \label{fig:Imp_rsp}
\end{figure}

Fig. \ref{fig:System_BD} illustrates the block diagram of the system. Given that the pulse generator sends an ideal delta function $\delta(t)$, the optical signal emitted by the LED has a bandwidth (BW) limited by $h_{LED} (t)$. Then, receiver $q$'th output,
$v^{(q)}(\boldsymbol\theta,t)$, can be written as 
\begin{equation}
    v^{(q)}(\boldsymbol\theta,t)= h_{LED} (t)*h_{ch}^{(q)} (\boldsymbol\theta,t)*h_{PD}^{(q)} (t) + n^{(q)}(t),	
    \label{rec_out}
\end{equation}
where $\boldsymbol\theta=(x,y)\in\mathbb{R}^2$ is the two-dimensional coordinate vector of the user's position to be estimated, $h_{ch}^{(q)} (\boldsymbol\theta,t)$ is the impulse response of the channel between the transmitter and the $q$’th PD, $h_{PD}^{(q)}(t)$ is the impulse response of the $q$’th PD and $n^{(q)}(t)$ is additive white Gaussian noise (AWGN). The system BW is defined as the bandwidth resulting from both $h_{LED} (t)$ and $h_{PD}^{(q)} (t)$. In order to model $h_{ch}^{(q)} (\boldsymbol\theta,t)$,  we employ the wireless optical channel model described in \cite{kahn1997wireless}. Fig. \ref{fig:Imp_rsp} illustrates $h_{ch}^{(q)} (\boldsymbol\theta,t)$ for 3 different user locations inside the room.


\begin{table}[h]
\centering
\caption{Simulation Parameters of Ray Tracing Channel Model \cite{lee2011indoor}.}
\label{TableI}
\begin{tabular}{|l|l|}
	\hline
	\textbf{Transmitter Parameters}         & \textbf{Value}  \\ \hline
	Height                                    & 0.85 m            \\ \hline
	Uplink wavelength                                & 950 nm            \\ \hline
	Lambertian mode (m)                       & 1                 \\ \hline
	LED transmit power, $P_T$                 & 10 mW             \\ \hline
	\textbf{Receiver Parameters}            & \textbf{Value}  \\ \hline
	Surface area of the PD, $A_{\text{PD}} $  & 1 $cm^ 2$         \\ \hline
	Height                                    & 3 m               \\ \hline
	Filed of view (Half Angle)                & $70^{\circ}$      \\ \hline
	PD1 location                          & ($150$ cm, $150$ cm, $300$ cm) \\ \hline 
    PD2 location                         & ($350$ cm, $150$ cm, $300$ cm)  \\ \hline
	PD3 location                          & ($150$ cm, $350$ cm, $300$ cm) \\ \hline 
    PD4 location                         & ($350$ cm, $350$ cm, $300$ cm)  \\ \hline
	\textbf{Room Parameters}                & \textbf{Value}  \\ \hline
	Room size (width $\times$ length $\times$ height)           & $5\times5\times3$ $m^3$  \\ \hline 
	Wall reflectance coefficient, $\rho$      & 0.8               \\ \hline
	Reflecting element area, $A_{\text{ref}}$ & $2\times 2$ cm$^2$              \\ \hline
\end{tabular}
\end{table}
\subsection{Fingerprinting}
To develop the proposed positioning algorithm, we first divide the indoor horizontal area into an $N \times M$ grid and then create a database of the channel impulse responses for different positions of the user on this grid, i.e. $C_k=(x_k,y_k), k \in \{1,2, \dots,MN\}$, for a known height $z$. In order to develop this fingerprinting map, we focus on strong features of the impulse response: the LOS peak power, $P_{LOS}$, the second power peak (SPP) term, $P_{SPP}$ (the height of the first peak of the diffuse term), and the time difference of arrival between these two components, $\Delta \tau$. The vector $S_k^{(q)}=[P_{k,LOS}^{(q)}, P_{k,SPP}^{(q)}, \Delta \tau_k^{(q)}], k \in \{1,2, \dots,MN\}$, represents receiver $q$th fingerprinting vector corresponding to the $k$th point on the measurement grid. All this data is assumed to be collected manually during an offline procedure, and crowd-sourcing methods can be developed to learn the fingerprint map automatically, as in \cite{7275499,8038447}.
\subsection{Encoder Design}
In order to estimate the location of one user, the user's transmitter has to send one narrow time pulse. However, to increase the SNR, a stationary transmitter could send a train of pulses with a large enough time period to avoid inter-symbol interference (ISI) at the receivers.

In a multiuser scenario, users have to send nearly orthogonal codes to make them distinguishable at the receivers. In this case, optical orthogonal codes (OOC) could be employed \cite{30982}. The length and weight of the OOC must be chosen in a way to minimize the ISI and inter-chip interference (ICI). Multi-pulse and multiuser performance is relegated to future studies.

\section{Positioning Algorithm}
\label{PA}
The first step of our localization algorithm is to combine features extracted from the different receivers. As we discuss above, in this work we consider at most three feature to be extracted from each receiver. Based on the number of  features used from each receiver, we refer to the algorithm as a one-feature algorithm (only extract the LOS feature from receivers), a two-feature algorithm (uses the LOS and SPP), and a three-feature algorithm (uses all three features, i.e., the LOS, SPP, and delay between the LOS and SPP). Applying a peak detector as a feature extractor on receiver $q$th output, $v^{(q)}(\boldsymbol\theta,t)$ in (\ref{rec_out}), the components of the observation vector for receiver $q$, $q=1, \cdots, Q$ (see Fig. \ref{fig:System_BD}) can be expressed as
\begin{align}
& v_{1}^{(q)}(\boldsymbol\theta)= P_{\text{LOS}}^{(q)}(\boldsymbol\theta)+ n^{(q)}_{1},\nonumber\\
& v_{2}^{(q)}(\boldsymbol\theta)= P_{\text{SPP}}^{(q)}(\boldsymbol\theta)+ n^{(q)}_{2},\nonumber\\
& v_{3}^{(q)}(\boldsymbol\theta)= \Delta \tau^{(q)}(\boldsymbol\theta)+ n^{(q)}_{3},
\label{CRB_EQ1}
\end{align}
where $Q$ is the number of receivers. $ n^{(q)}_{1}$ and $n^{(q)}_{2}$ are zero mean independent Gaussian noises with variance $\sigma^2$. $n_{3}$ is the time delay noise between LOS and SPP, which is modeled as a zero mean Gaussian noise with variance $\sigma_{\tau}^2$  and assumed independent from $ n^{(q)}_{1}$ and $n^{(q)}_{2}$. Let $\boldsymbol V^{(q)}(\boldsymbol\theta)=[v_1^{(q)}(\boldsymbol\theta), v_2^{(q)}(\boldsymbol\theta), v_3^{(q)}(\boldsymbol\theta)]$ and $\boldsymbol n^{(q)}=[n_1^{(q)}, n_2^{(q)}, n_3^{(q)}]$ be the observation vector and noise vector at receiver $q$, respectively. Then, the observation  and noise vectors are concatenated into supervectors as $\boldsymbol V(\boldsymbol\theta)=[\boldsymbol V^{(1)}(\boldsymbol\theta),\boldsymbol V^{(2)}(\boldsymbol\theta),\cdots,\boldsymbol V^{(Q)}(\boldsymbol\theta)]^T$ and $\boldsymbol n=[\boldsymbol n^{(1)},\boldsymbol n^{(2)},\cdots, \boldsymbol n^{(Q)}]$, respectively. 
The overall measurement covariance matrix is 
$\boldsymbol\Sigma=COV(\boldsymbol n)=diag(\sigma^2,\sigma^2,\sigma_{\tau}^2,\cdots,\sigma^2,\sigma^2,\sigma_{\tau}^2)$.

\begin{figure}
	\centering
	\begin{subfigure}{}
		\includegraphics[width=3.2in]{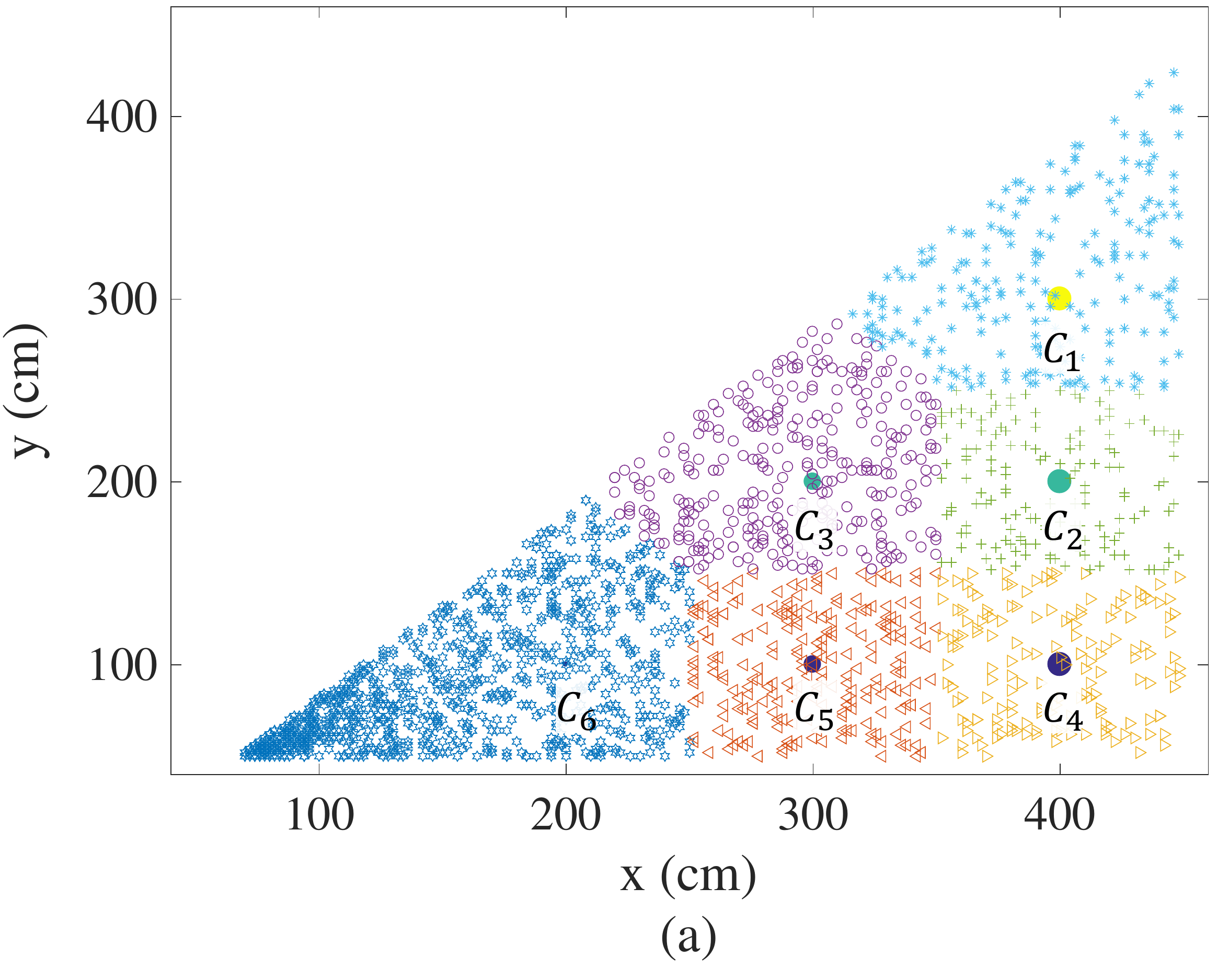}
		\label{fig:Room}
	\end{subfigure}
	\begin{subfigure}{}
		\includegraphics[width=3.2in]{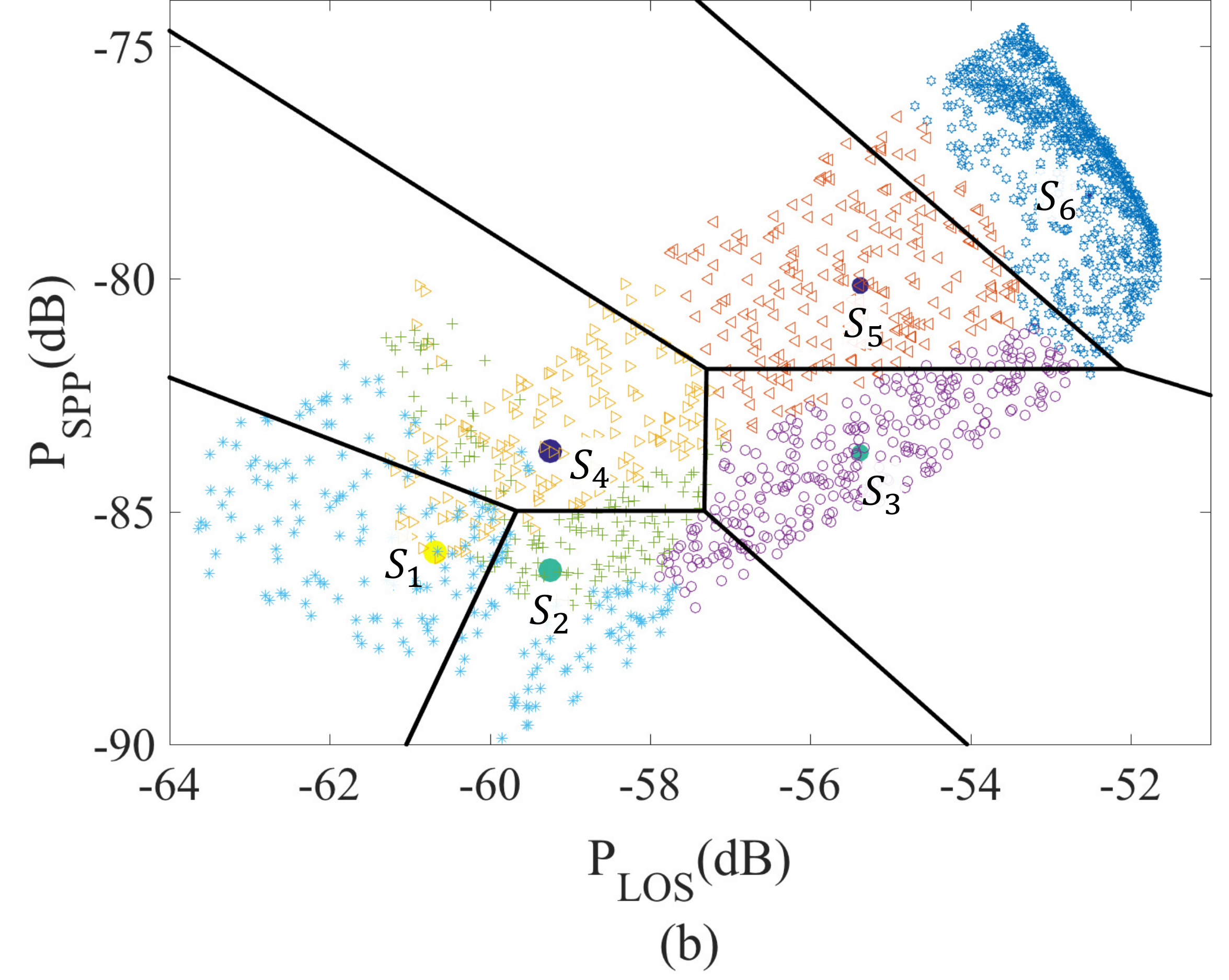} 
		\label{fig:Decision_reg}
	\end{subfigure}   
	\caption{Mapping the room plane onto the observation plane. a) room cells distinguished by color b) Corresponding features and nearest-neighbor decision regions. Random points in both planes are denoted with corresponding markers of the same size and color.}
	\label{fig:Decision_reg_Room} 
\end{figure}

In order to locate the user, we employ the nearest neighbor algorithm to find the grid point $C_k=(x_k,y_k)$ corresponding to the fingerprinting supervector nearest the received features in the Euclidean sense, defined as $\boldsymbol{S_{\hat{k}}}=[S_{\hat{k}}^{(1)}, S_{\hat{k}}^{(2)}, \cdots, S_{\hat{k}}^{(Q)}]^T$. The index of the selected grid point, $\hat{k}$, can be obtained as
\begin{align}
\hat{k} =& \ \begin{matrix} \arg \min\limits_{k}
\left (  \boldsymbol V(\boldsymbol\theta)-\boldsymbol S_{k}\right ) {\boldsymbol\Sigma}^{-1}\left (  \boldsymbol V(\boldsymbol\theta)-\boldsymbol S_{k}\right )^T,
\end{matrix} \nonumber \\& k \in \{1,2, \dots,NM\}
\label{error_drivation2} 
\end{align}
Fig. \ref{fig:Decision_reg_Room}-(a), and (b) illustrate an example of grid points $C_k=(x_k,y_k)$ in the room plane and the corresponding constellation points $\boldsymbol{S_{k}}$ in the observation plane, based on two observations components, $P_{\text{LOS}}$ and $P_{SPP}$, for one PD. Considering no AWGN noise, the distribution of observation features related to each grid cell is illustrated in \ref{fig:Decision_reg_Room}-(b). 
In addition, the nearest neighbor decision regions corresponding to grid cells within the room given in Fig. \ref{fig:Decision_reg_Room}-(a) are shown in Fig. \ref{fig:Decision_reg_Room}-(b).

There exist three different sources of localization error for the proposed algorithm. One error source is the AWGN noise that is introduced in (\ref{CRB_EQ1}). The second error source arises when the user is not located exactly on the grid points. No matter how accurately the algorithm can find the closest grid point, there is always an offset error between the real location and the closest grid point in the room, which is known as quantization error. If the transmitter is randomly positioned in a grid cell of size ${\Delta}^2$ according to a uniform distribution, the RMS quantization error is equal to ${\Delta }/{\sqrt{6}}$, where $\Delta $ is the grid step size. We refer to this as the quantization lower bound (QLB).
The third error source is due to the fact that the impulse response features observed within the entire grid cell corresponding to grid point $C_k=(x_k,y_k)$ are not equal to those of the corresponding detection constellation points $\boldsymbol S_{k}$ (see Fig. \ref{fig:Decision_reg_Room}-b).

The root mean square (RMS) positioning error can be calculated as \cite{7996815}
\begin{equation} 
d_{RMS}(\mathbf{\boldsymbol\theta})=\sqrt{\mathbf{E}_{\boldsymbol\theta}\left \{ \sum_{i=1}^{N\times M}\left | \boldsymbol\theta-\boldsymbol C_i \right |^2 \epsilon_i({\boldsymbol\theta}) \right \}}
\label{error_drivation3} 
\end{equation}
where $\epsilon_i({\boldsymbol\theta})$ is the probability the positioning algorithm chooses the grid point $C_i$, given the user is located at position $\mathbf{\boldsymbol\theta}$. $\mathbf{E_{\boldsymbol\theta}}$ denotes the expectation conditioned on $\boldsymbol\theta $.

\begin{figure}[!t]
	\centering
	\includegraphics[width=3.2in]{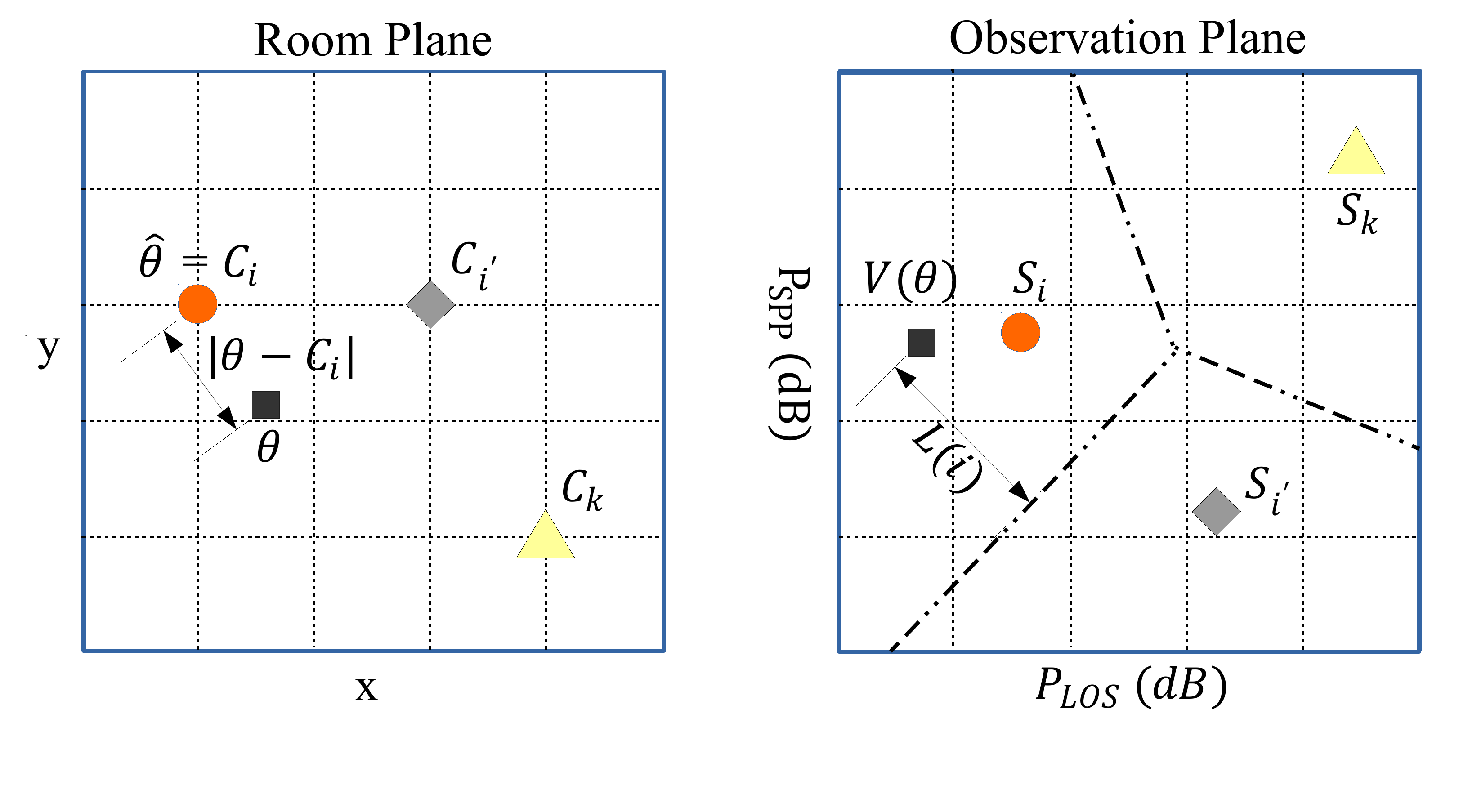} 
	\caption{Mapping the room plane onto the observation plane. The $S_i$, $S_{i'}$, and $S_k$ are three typical observation points, close together in observation space and, the $C_i$, $C_{i'}$, and $C_k$ are corresponding center points in the room plane. }
	\label{fig:Mapping_scatch1}
\end{figure}

 Due to the complexity of the decision regions shapes, a derivation of the exact value of $\epsilon_i({\boldsymbol\theta})$ is nontrivial. Hence, a practical approach is to calculate a lower bound (LB) on the estimation error that is tight in high SNR scenarios. In this regard, we consider only $i$'s in (\ref{error_drivation3}) that correspond to the two closest center points to the observation point, $\boldsymbol V(\boldsymbol\theta)$. Then,  $\epsilon_i({\boldsymbol\theta})$ evaluated at these center points can be written as
\begin{align}
\epsilon_i({\boldsymbol\theta})=&1- Q(\sqrt {\frac{{{\textbf{L}^T}(i)}\cdot{\boldsymbol{\Sigma}}^{-1}\cdot{\textbf{L}}(i)}{2}}),
\end{align}
and $\epsilon_{i'}({\boldsymbol\theta})=1-\epsilon_i({\boldsymbol\theta})$, where ${\textbf{L}}(i)$ is the distance vector between $\boldsymbol V(\boldsymbol\theta)$ and the boundary between the two closest center points, named $i$ and $i'$ (see Fig.~\ref{fig:Mapping_scatch1}).
\section{Cramer Rao Lower Bound (CRLB) of the Positioning Error}
\label{Sec:CRLB}
In this section, we develop the CRLB for any unbiased estimator of a user's position based on the observation vector $\boldsymbol V(\boldsymbol\theta)$, in additive Gaussian noise, for a multiple PDs scenario. As described above, our algorithm considers a quantized fingerprinting map. However, in order to calculate the ultimate estimation accuracy for the proposed localization algorithm, in this section, we consider the fingerprint map as a continuous surface in the calculation of the CRLB. The quantization effect can simply be added to the CRLB as the two sources of errors are independent.

Given the fact that all the noise terms in (\ref{CRB_EQ1}) are assumed to be Gaussian distributed, the joint probability density function of the observation vector, $\boldsymbol V(\boldsymbol\theta)$, conditioned on $\boldsymbol\theta$ is a multivariate Gaussian distribution,

\begin{equation}
f(\boldsymbol V(\boldsymbol\theta)|{\boldsymbol\theta})=\frac{\exp(-\frac{1}{2}\boldsymbol V(\boldsymbol\theta)\boldsymbol\Sigma^{-1}\boldsymbol V^T(\boldsymbol\theta))}{(\sqrt{2\pi})^{3Q}\left | \mathbf{\Sigma}  \right |}
\label{CRB_EQ2}
\end{equation}
The Fisher information matrix (FIM), $ \mathbf{J}(\boldsymbol\theta)$, is defined as \cite{scharf1991statistical} 
\begin{equation}
\mathbf{J}(\mathbf{\theta} )=-\mathbf{E_{\boldsymbol\theta}}{\frac{\partial }{\partial {\boldsymbol\theta} }\left(\frac{\partial}{\partial {\boldsymbol\theta}} \ln \left [f(\boldsymbol V(\boldsymbol\theta)|{\boldsymbol\theta})  \right ] \right )^T}=\boldsymbol{\mathbf{H}}\ \mathbf{J(V)}\ {\mathbf{H^T}}
\label{CRB_EQ4}
\end{equation}
where $ \mathbf {J}(\mathbf{V})=-\mathbf{E}_{V}\frac{\partial }{\partial {\mathbf{V}} }\left(\frac{\partial}{\partial {\mathbf{V}}} \ln \left [f(\boldsymbol V(\boldsymbol\theta)|{\boldsymbol\theta})  \right ] \right )^T$. Based on (\ref{CRB_EQ2}), $\mathbf {J}(\mathbf{V})$ can be expressed as $\mathbf{J}(\mathbf{V})=\boldsymbol\Sigma^{-1}$. The matrix $\boldsymbol H$ is defined as
\begin{align}
\boldsymbol H=\mathbf{E}\begin{pmatrix}
\frac{\partial v_{1}^{(1)}}{\partial {\theta}_1} &  \frac{\partial v_{2}^{(1)}}{\partial {\theta}_1} & \frac{\partial v_{3}^{(1)}}{\partial {\theta}_1} & \dots & \frac{\partial v_{1}^{(Q)}}{\partial {\theta}_1} &  \frac{\partial v_{2}^{(Q)}}{\partial {\theta}_1} & \frac{\partial v_{3}^{(Q)}}{\partial {\theta}_1}\\ 
\frac{\partial v_{1}^{(1)}}{\partial {\theta}_2}  & \frac{\partial v_{2}^{(1)}}{\partial {\theta}_2}  & \frac{\partial v_{3}^{(1)}}{\partial {\theta}_2} & \dots & \frac{\partial v_{1}^{(Q)}}{\partial {\theta}_2}  & \frac{\partial v_{2}^{(Q)}}{\partial {\theta}_2}  & \frac{\partial v_{3}^{(Q)}}{\partial {\theta}_2}
\end{pmatrix}
\label{CRB_EQ05}
\end{align}
We derive the components of $\boldsymbol H$ based on the channel model described in \cite{kahn1997wireless}. Given the Lambertian equations for the LOS component, $\mathbf{E}\{v_{1}^{(q)}\}=P_{LOS}^{(q)}(\boldsymbol\theta)$ can be written as
\begin{align}
&\mathbf{E}\{v_{1}^{(q)}\}= P_{LOS}^{(q)}( \theta_{1}, \theta_{2})=\frac{m+1}{2\pi d_q^{2}}A_R^{(q)} \cos^m\left ( \Phi^{(q)}  \right )\cos\left ( \Psi^{(q)}  \right )\nonumber \\
&=\frac{\left (m+1  \right )A_R^{(q)} z^{m+1}}{2\pi \left (\sqrt{z^2+\left ( \theta_{1}-x_{r,q} \right )^2+\left (  \theta_{2}-y_{r,q} \right )^2}  \right )^{m+3}},
\label{CRB_EQ8}
\end{align}
where $A_R^{(q)}$ is the surface area of the $q$th PD, $\Phi^{(q)}$ is the irradiance angle, $\Psi^{(q)}$ is the angle of incidence with respect to the $q$th receiver axis. $(x_{r,q},y_{r,q})$ is the $q$th receiver coordinates, and $d_q$ is the distance between the user and receiver $q$. $m$ is the order of Lambertian emission of the transmitter LED.

The first partial derivative of the LOS component can be obtained as
\begin{align}
&\mathbf{E}\left \{\frac{\partial v_{1}^{(q)}}{\partial  \theta_{1}}  \right \}=\frac{\partial }{\partial  \theta_{1}}P_{LOS}^{(q)}(\boldsymbol\theta)=-G_0\frac{ \left (  \theta_{1}-x_{r,q} \right )}{d_q^{m+4}},\nonumber \\
&\mathbf{E}\left \{\frac{\partial v_{1}^{(q)}}{\partial  \theta_{2}}  \right \}=\frac{\partial }{\partial  \theta_{2}}P_{LOS}^{(q)}(\boldsymbol\theta)=-G_0\frac{ \left (  \theta_{2}-y_{r,q}\right )}{d_q^{m+4}},
\label{CRB_EQ9}
\end{align}
where $G_0$ is defined as $G_0=\frac{(m+1)\left ( m+3 \right )z^{m+1}A_R}{2 \pi}$.
The derivatives of the $\mathbf{E}\{v_{2}^{(q)}\}=P_{SPP}^{(q)}( \theta_{1}, \theta_{2})$ and $\mathbf{E}\{v_{3}^{(q)}\}=\Delta \tau^{(q)}(\theta_{1}, \theta_{2})$ components can be approximated using quadratic regression
\begin{align}
& \mathbf{E}\{v_{2}^{(q)}\}=P_{SPP}^{(q)}(\boldsymbol\theta)=\mathbf{A}(\theta_{1})^T \mathbf{U}^{(q)} \mathbf{A}(\theta_{2}),\nonumber \\
& \mathbf{E}\{v_{3}^{(q)}\}=\Delta \tau^{(q)}(\boldsymbol\theta)=\mathbf{A}(\theta_{1})^T \mathbf{T}^{(q)} \mathbf{A}(\theta_{2}), 
\label{CRB_apx11}
\end{align}
where $\mathbf{A}(x)=\begin{bmatrix} 1 \ x \ x^2 \ x^3 \ x^4 \end{bmatrix}^T$. $\mathbf{U}^{(q)}$ and $\mathbf{T}^{(q)}$ depend on the geometry of the room, as explained in the Appendix. The first partial derivative of the SPP and $\Delta \tau$ components can be written as
\begin{align}
& \mathbf{E}\left \{\frac{\partial v_{2}^{(q)}}{\partial  \theta_{1}}  \right \}=\frac{\partial }{\partial  \theta_{1}}P_{SPP}^{(q)}(\boldsymbol\theta)=\mathbf{\dot{A}}(\theta_{1})^T \mathbf{U}^{(q)} \mathbf{A}(\theta_{2}), \nonumber \\
& \mathbf{E}\left \{\frac{\partial v_{2}^{(q)}}{\partial  \theta_{2}}  \right \}=\frac{\partial }{\partial  \theta_{2}}P_{SPP}^{(q)}(\boldsymbol\theta)=\mathbf{A}(\theta_{1})^T \mathbf{U}^{(q)} \dot{\mathbf{A}}(\theta_{2}),
\label{CRB_EQ10}
\end{align}
where $\mathbf{\dot{A}}(x)=\begin{bmatrix} 0 \ 1 \ 2x \ 3x^2 \ 4x^3 \end{bmatrix}^T$. Similarly,
\begin{align}
& \mathbf{E}\left \{\frac{\partial v_{3}^{(q)}}{\partial  \theta_{1}}  \right \}=\frac{\partial }{\partial  \theta_{1}}\Delta \tau^{(q)}(\boldsymbol\theta)=\mathbf{\dot{A}}(\theta_{1})^T \mathbf{T}^{(q)} \mathbf{A}(\theta_{2}), \nonumber \\
& \mathbf{E}\left \{\frac{\partial v_{3}^{(q)}}{\partial  \theta_{2}}  \right \}=\frac{\partial }{\partial  \theta_{2}}\Delta \tau^{(q)} (\boldsymbol\theta)=\mathbf{A}(\theta_{1})^T \mathbf{T}^{(q)} {\dot{\mathbf{A}}(\theta_{2})}.
\label{CRB_EQ12}
\end{align}
Substituting (\ref{CRB_EQ12}), (\ref{CRB_EQ10}), and (\ref{CRB_EQ9}) in (\ref{CRB_EQ05}), $\boldsymbol H$ is obtained, and based on (\ref{CRB_EQ4}), the components of the $2\times2$ matrix $\mathbf{J}(\mathbf{\theta} )$ can be found as
\begin{align*}
& J_{11}(\mathbf{\boldsymbol\theta} )= \sum_{q=1}^{Q}G_0^2\frac{\left ( m+3 \right )^2\left (  \theta_{1}-x_{r,q} \right )^2}{\sigma^2d_q^{m+4}}\nonumber\\ +&\frac{({\mathbf{\dot{A}}(\theta_{1})^T \cdot \mathbf{U}^{(q)}\cdot \mathbf{A}(\theta_{2})})^2}{\sigma^2} 
+\frac{({\mathbf{\dot{A}}(\theta_{1})^T \cdot \mathbf{T}^{(q)}\cdot \mathbf{A}(\theta_{2})})^2}{\sigma_T^2}
\end{align*}
\begin{align*}
& J_{12}(\mathbf{\boldsymbol\theta} )=J_{21}(\mathbf{\boldsymbol\theta} )= \sum_{q=1}^{Q}G_0^2\frac{\left ( m+3 \right )^2\left (  \theta_{1}-x_{r,q} \right )\left (  \theta_{2}-y_{r,q} \right )}{\sigma^2d_q^{m+4}}\nonumber\\ &+\frac{({\mathbf{\dot{A}}(\theta_{1})^T \cdot \mathbf{U}^{(q)}\cdot \mathbf{A}(\theta_{2})})({\mathbf{A}(\theta_{1})^T \cdot \mathbf{U}^{(q)}\cdot \mathbf{\dot{A}}(\theta_{2})})}{\sigma^2} \nonumber\\
&+\frac{({\mathbf{\dot{A}}(\theta_{1})^T \cdot \mathbf{T}^{(q)}\cdot \mathbf{A}(\theta_{2})})({\mathbf{A}(\theta_{1})^T \cdot \mathbf{T}^{(q)}\cdot \mathbf{\dot{A}}(\theta_{2})})}{\sigma_T^2}
\end{align*}
\begin{align*}
 J_{22}(\mathbf{\boldsymbol\theta} )= \sum_{q=1}^{Q}G_0^2\frac{\left ( m+3 \right )^2\left (\theta_{2}-y_{r,q} \right )^2}{\sigma^2d_q^{m+4}}
\end{align*}
\begin{align*}
+\frac{({\mathbf{A}(\theta_{1})^T \cdot \mathbf{U}^{(q)}\cdot \mathbf{\dot{A}}(\theta_{2})})^2}{\sigma^2} 
+\frac{(\mathbf{A}{(\theta_{1})^T \cdot \mathbf{T}^{(q)}\cdot \mathbf{\dot{A}}(\theta_{2})})^2}{\sigma_T^2}
\end{align*}
\section{Numerical Results and Discussion}
\label{Sec:Mum}
In this section, we present numerical results of the performance of the proposed localization algorithm. The channel modeling based on ray tracing and the system configuration are discussed first, followed by a description of the Monte-Carlo simulation. We then explain the behavior of the system measured by the RMS estimation error as a function of SNR, grid step size, and transmitter BW, for a different number of PDs. We compare the LB and Monte-Carlo results for the first two parameters, i.e., SNR and grid step size. Finally, the CRLB results are discussed, as a function of SNR.

The ray tracing channel model parameters are listed in Table \ref{TableI}. We consider the system configuration illustrated in Fig. 1 with at most $Q=4$ PDs on the ceiling. The transmitter is assumed to be at a fixed known height from the floor facing vertically upwards.

\subsection{Estimation Performance for Infinite BW Assumption}
In order to determine the estimation error, we run a Monte Carlo (MC) simulation by randomly choosing the location of the user and estimating the closest grid point on the room plane using the proposed algorithm. The RMS positioning error is a result of choosing a constellation point by the algorithm and the mapping cost of that decision. However, given that the algorithm tries to find the closest grid point, the larger the grid step size, the worse the absolute positioning accuracy becomes. And yet increasing the grid step size lowers the probability of mapping to some far bin in the room. Therefore, there is a trade-off between the room plane absolute grid accuracy and the cost of wrong mapping.

\begin{figure}
 \centering
 \begin{subfigure}{}
 	 \hspace{-0.1in}
  \includegraphics[width=3.2in]{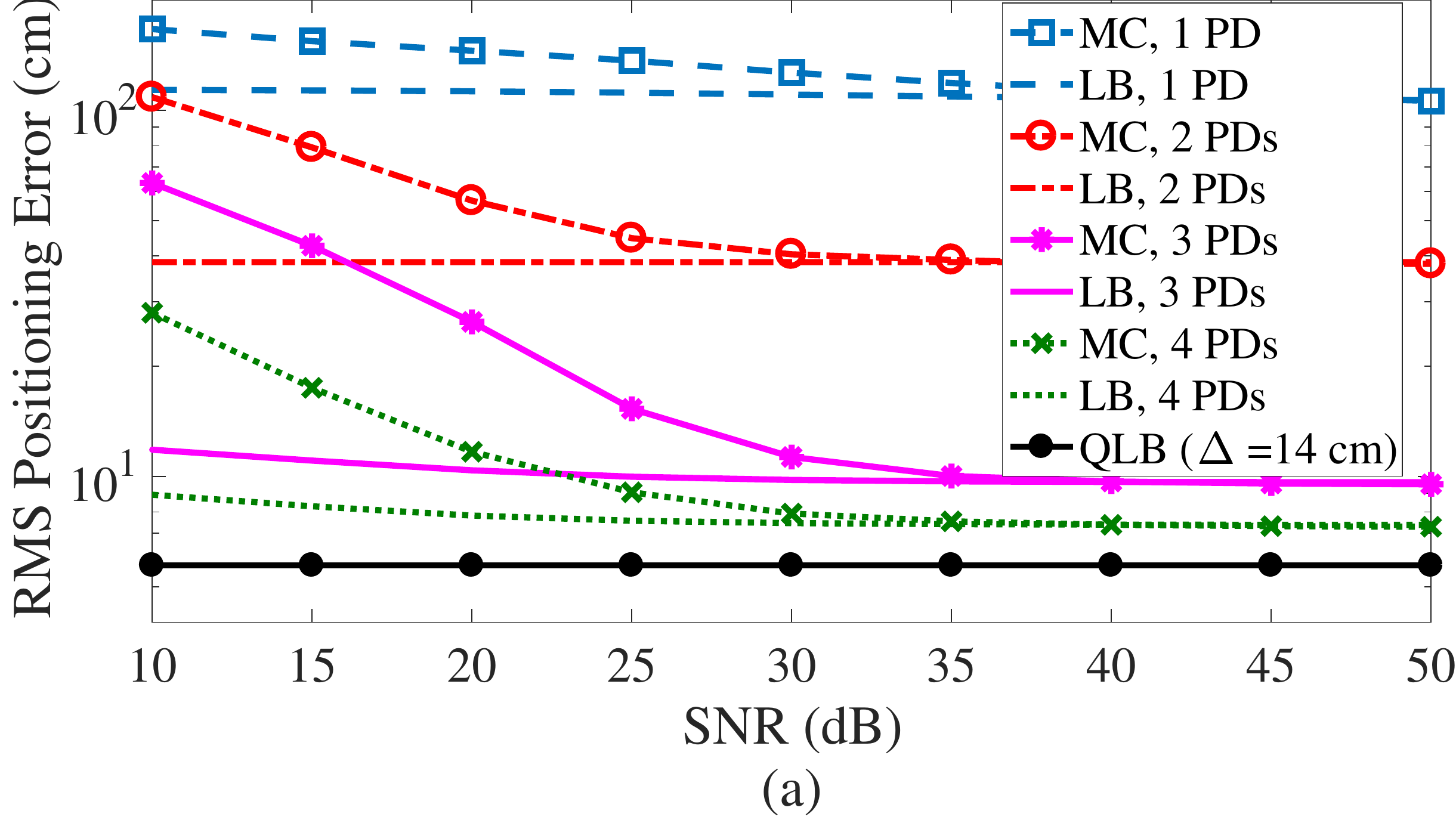}
  \label{fig:RMSSNRa}
 \end{subfigure}
 \begin{subfigure}{}
  \includegraphics[width=3.2in]{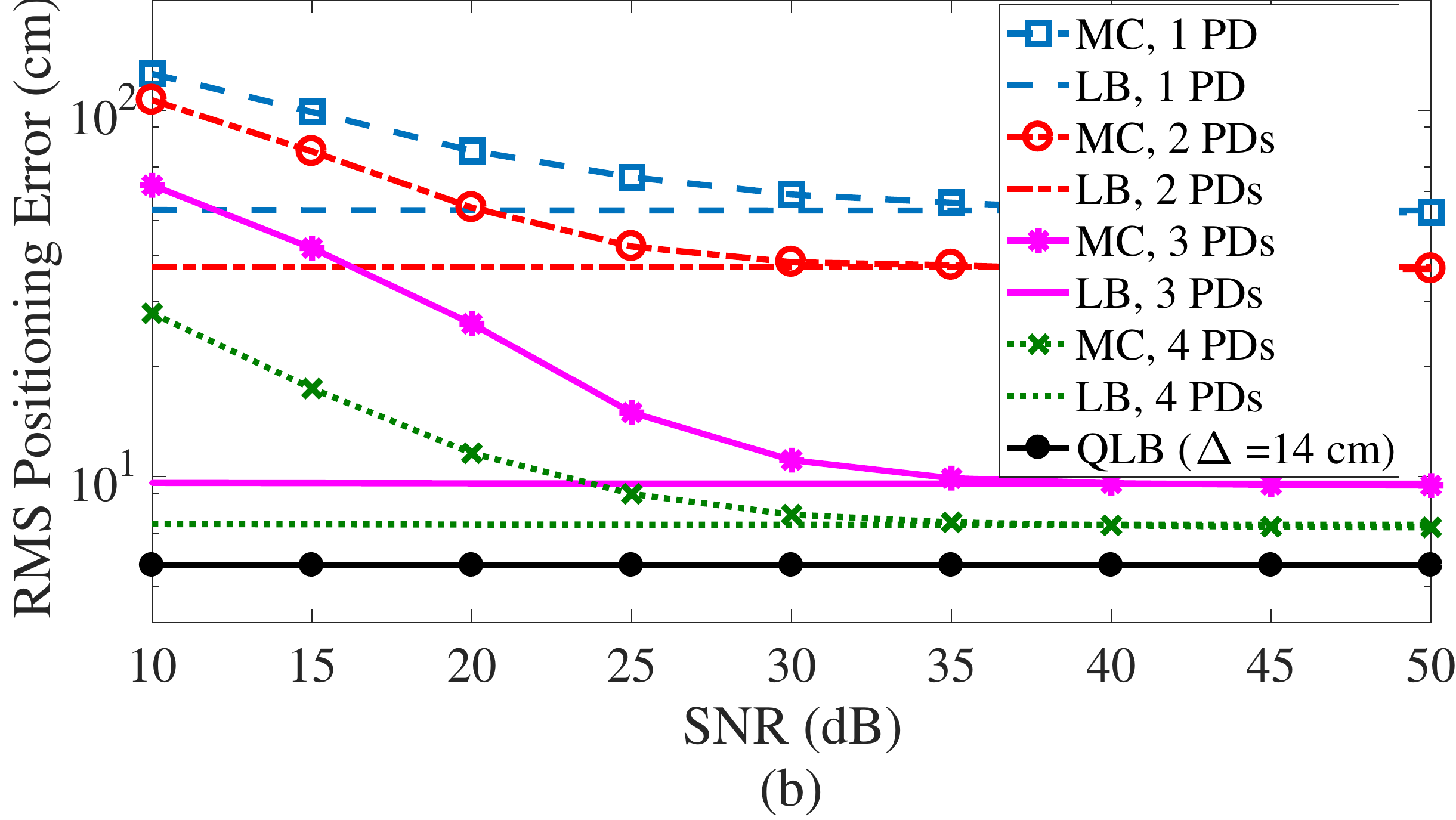} 
  \label{fig:RMSSNRb}
 \end{subfigure}   
 \caption{RMS distance error for a grid step size of $14$~cm and multiple PDs scenarios:
 a) two features, and b) three features.}
 \label{fig:RMSSNR} 
\end{figure}

Figs.~\ref{fig:RMSSNR}-(a), (b) show the RMS positioning errors for the two and three features algorithms, respectively. For all cases considering different numbers of PDs, the simulated error converges to the LB results (calculated in Section \ref{PA}) at high SNR, which verifies the analytical results. For the one PD case, we need 3 features to attain a $25$ cm accuracy. Deploying more PDs leads to a higher accuracy: for 4 PDs we reach the minimum error of QLB= $5.7$ cm (defined in Section \ref{PA}) at an SNR of $30$ dB. This error is inevitable due to  quantization based on the grid. As expected, the 3 features algorithm outperforms the  2 features algorithm for the same SNR and number of PDs, especially for a smaller number of PDs. However, both algorithms reach the best accuracy, the QLB, at high SNR and a sufficient number of PDs.

\begin{figure}[h]
	\centering
	\includegraphics[width=3.3in]{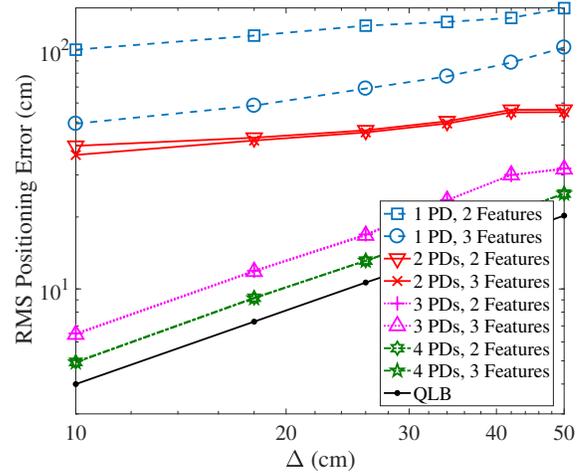} 
	\caption{RMS distance error for SNR$=30~\si{dB}$ and different number of PDs and features as the grid step size increases.}
	\label{fig:RMS_Grid} 
\end{figure}
Fig. \ref{fig:RMS_Grid} shows the RMS positioning error as a function of grid step size for a high SNR scenario. 
There is a nearly linear relation between accuracy and the grid step size for all multiple PDs scenarios. In addition, for a larger number of PDs and in the high SNR case, the RMS error gets closer to the RMS QLB, where the estimated location is mapped to the closest grid point in the room.
\subsection{CRLB as a Function of SNR}
\begin{figure}
	\centering
		\begin{subfigure}{}
		\includegraphics[width=3.3in]{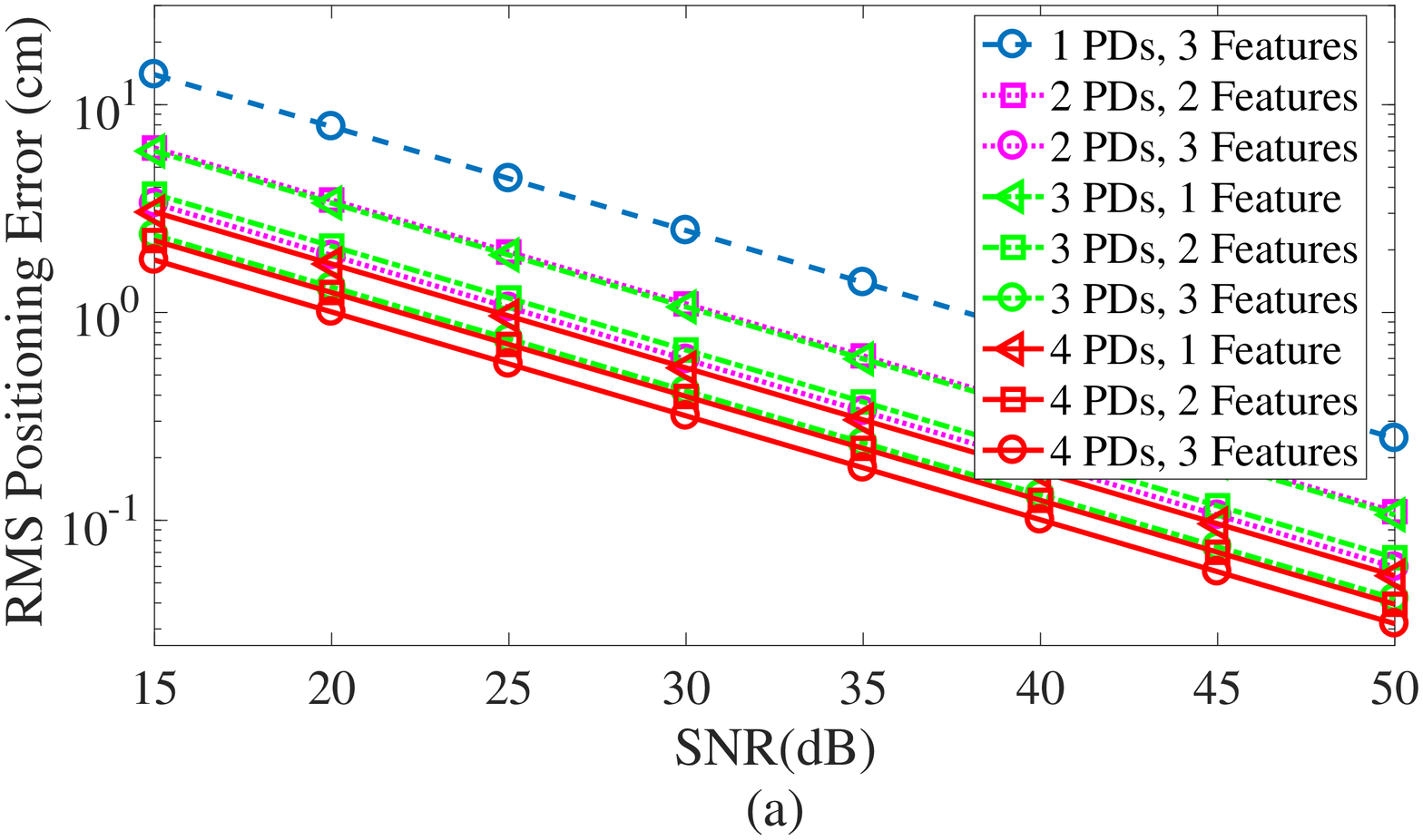}
	\end{subfigure}
	~ 
	\begin{subfigure}{}
		\includegraphics[width=3.3in]{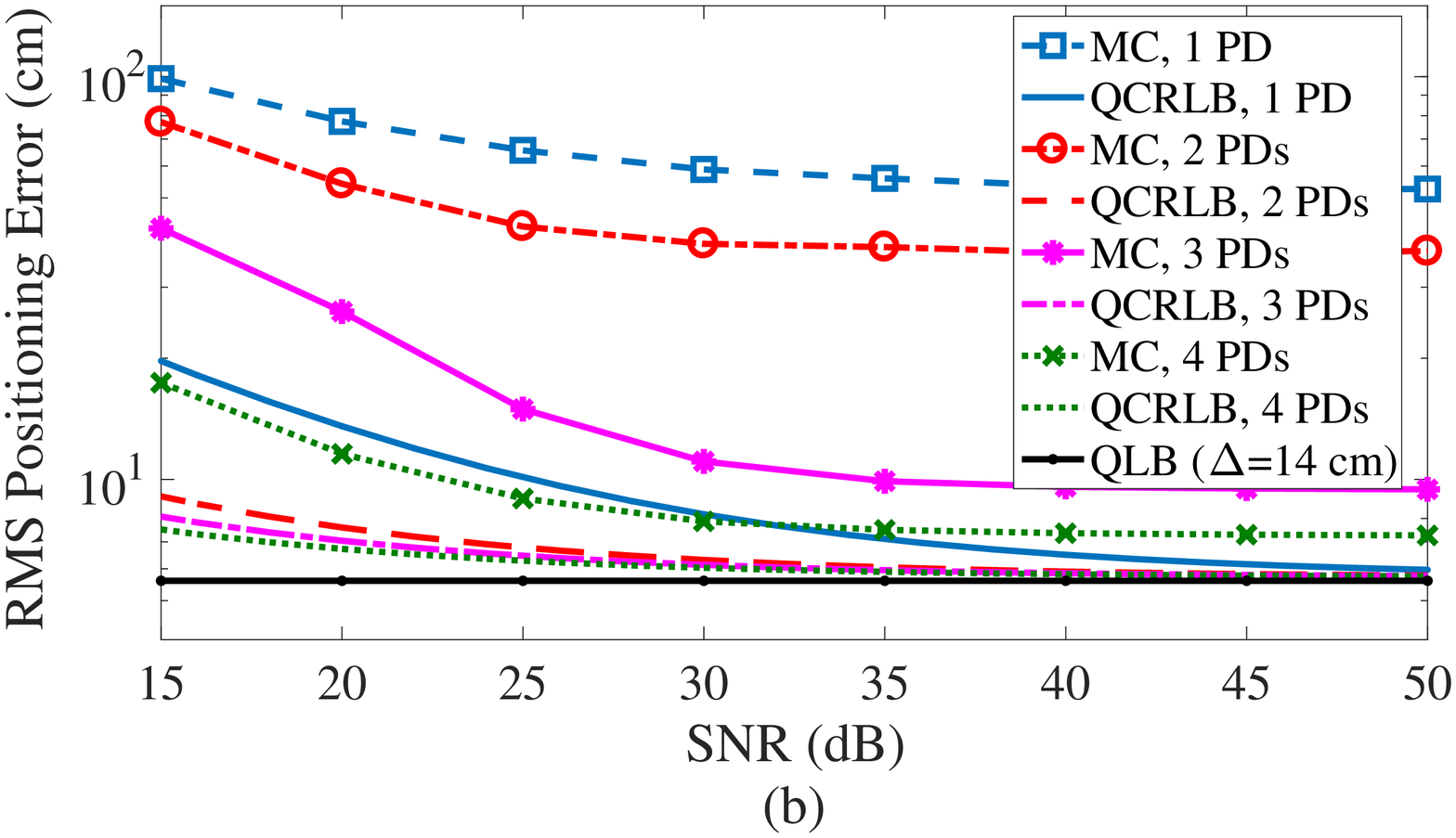} 
	\end{subfigure}   
	\caption{RMS distance error for a grid step size of $14$~cm and multiple PDs scenarios. a) CRLB for different number of PDs and features, b) comparison of QCRLB and MC for three features.}
	\label{fig:CRLB} 
\end{figure}

Fig. \ref{fig:CRLB}-a illustrates the CRLB for different numbers of PDs. The estimation accuracy as measured by the RMS error increases by one order of magnitude per $20$ dB of SNR increment. For the same number of features, our algorithm with 2 PDs, and 2 features (LOS and SPP) outperforms the basic RSS method (explained in \cite{yang2013indoor}) with 4 PDs, one-feature component (just the LOS). For the same number of PDs, increasing the number of features employed leads to more accurate estimation.

Given the fact that the additive noise and quantization error are independent, the quantization effect on the CRLB can be obtained by adding CRLB and QLB variances; we name this the quantized Cramer-Rao lower bound (QCRLB). Fig. \ref{fig:CRLB}-b shows MC results compared with the corresponding QCRLB for a three features scenario and grid step size equal to $14$ cm. The MC results get close to the QCRLB at SNR higher than $30$ dB, for the 3 and 4 PDs scenarios. For other cases, the positioning error is dominated by the probability of choosing the wrong cell as the nearest-neighbor in the observation plane.

\subsection{RMS Error as a Function of Bandwidth}
\begin{figure}[h]
	\centering
	\includegraphics[width=3in]{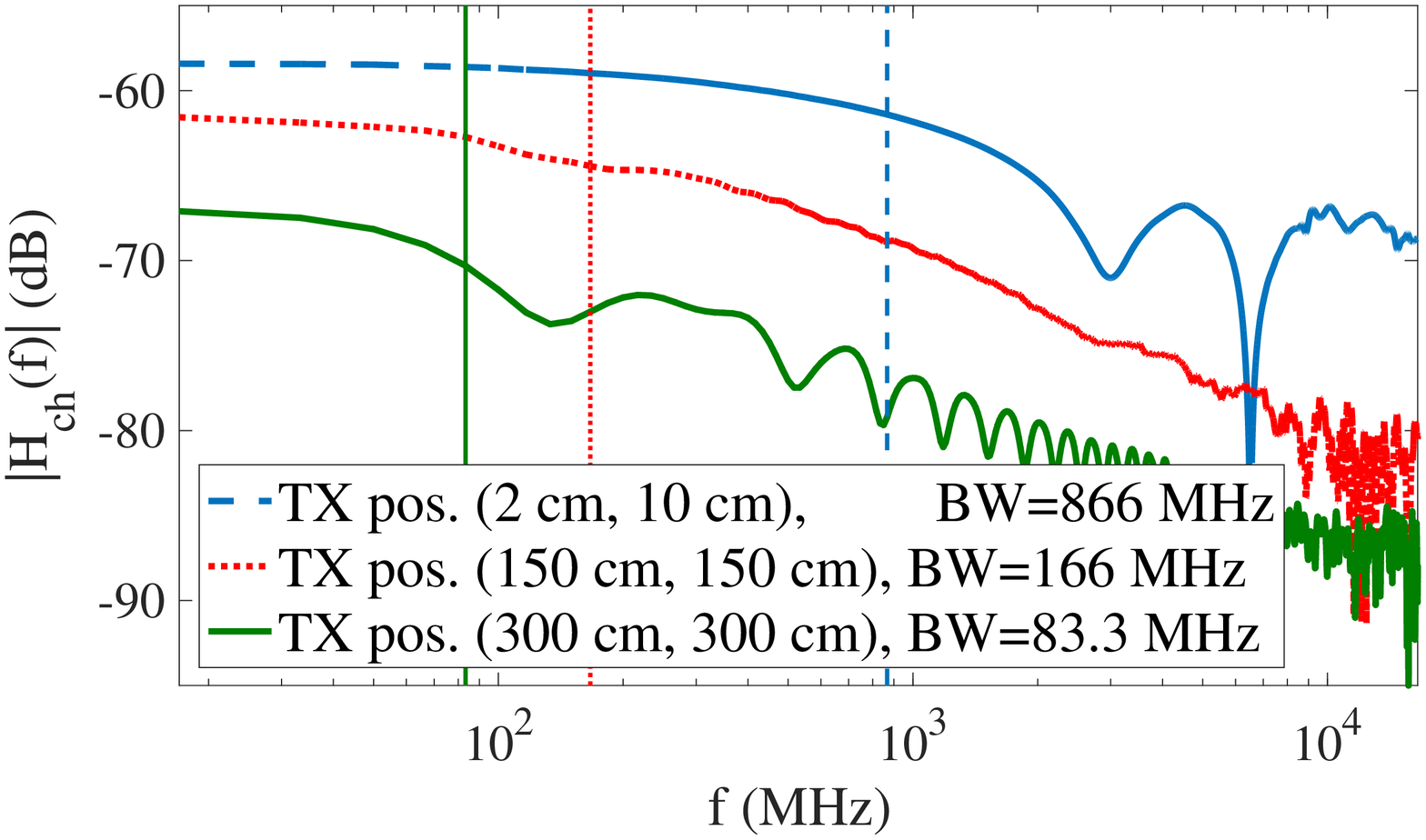} 
	\caption{The frequency response of the channel $H_{ch}(\theta,f)$ between  the transmitter located at different locations in the room and the receiver located at PD1 position (see Table \ref{TableI}).}
	\label{fig:FreqRespFDL}
\end{figure}
\begin{figure}[h]
	\centering
	\includegraphics[width=3.3in]{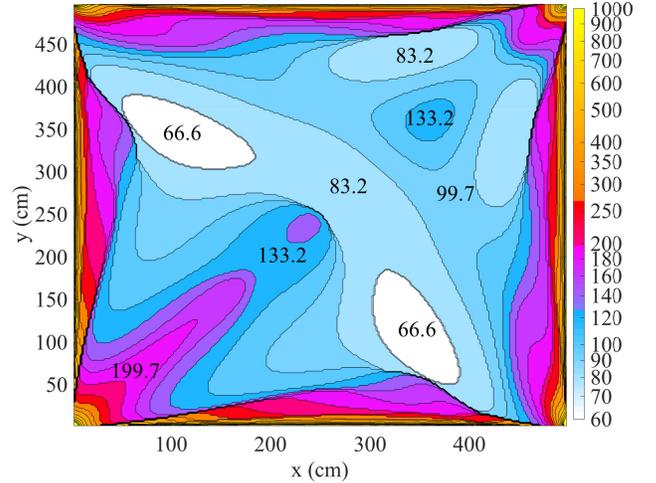} 
	\caption{Contours of $3$ dB channel BW for different locations of the transmitter in the room and the receiver located at the orange dot, which is the PD1 position in Table \ref{TableI}.}
	\label{fig:BW_Surf}
\end{figure}
\begin{figure}
	\centering
	\begin{subfigure}{}
		\includegraphics[width=3.2in]{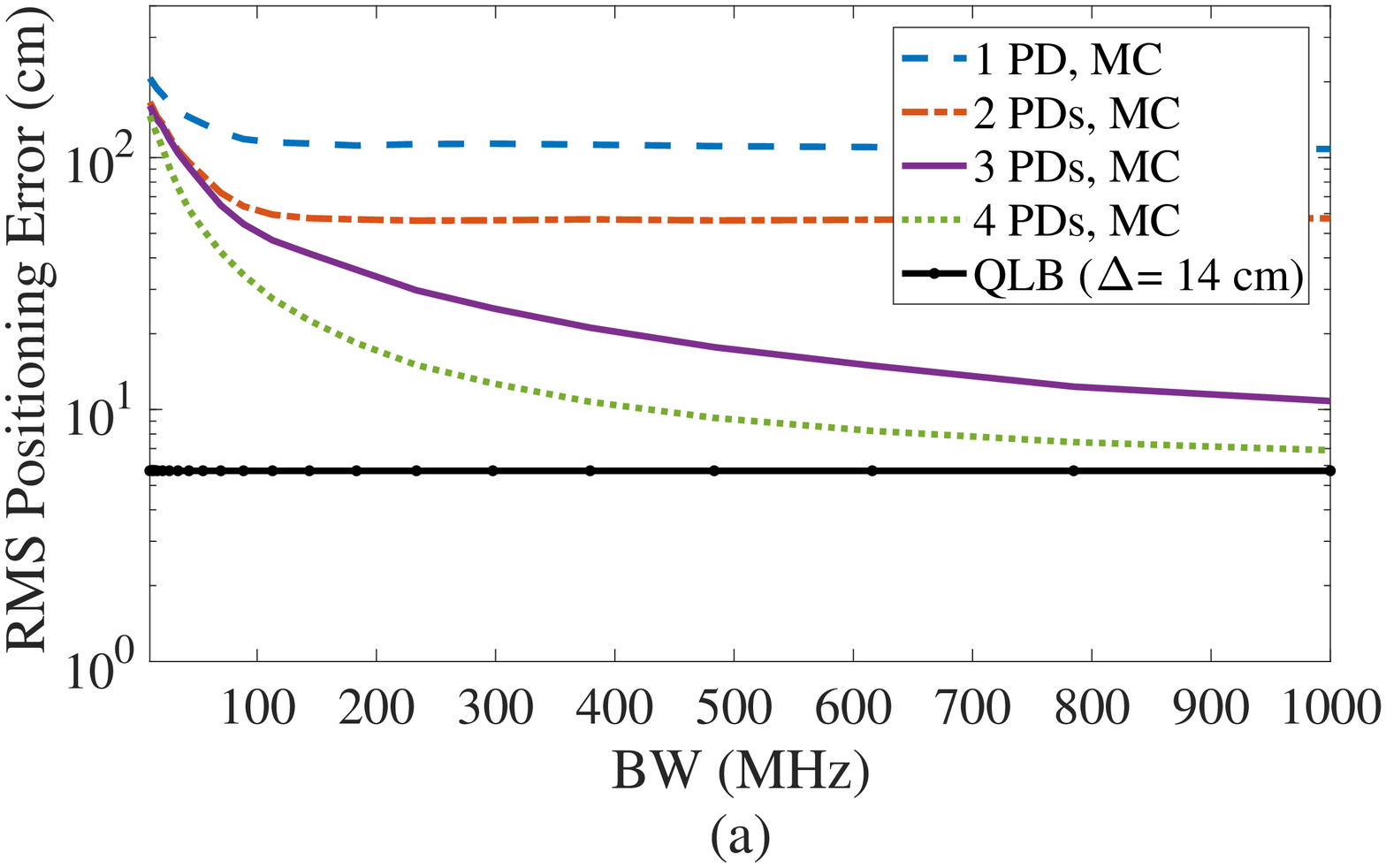}
		\label{fig:BW_2obs}
	\end{subfigure}
	\vfill
	\begin{subfigure}{}
		\includegraphics[width=3.2in]{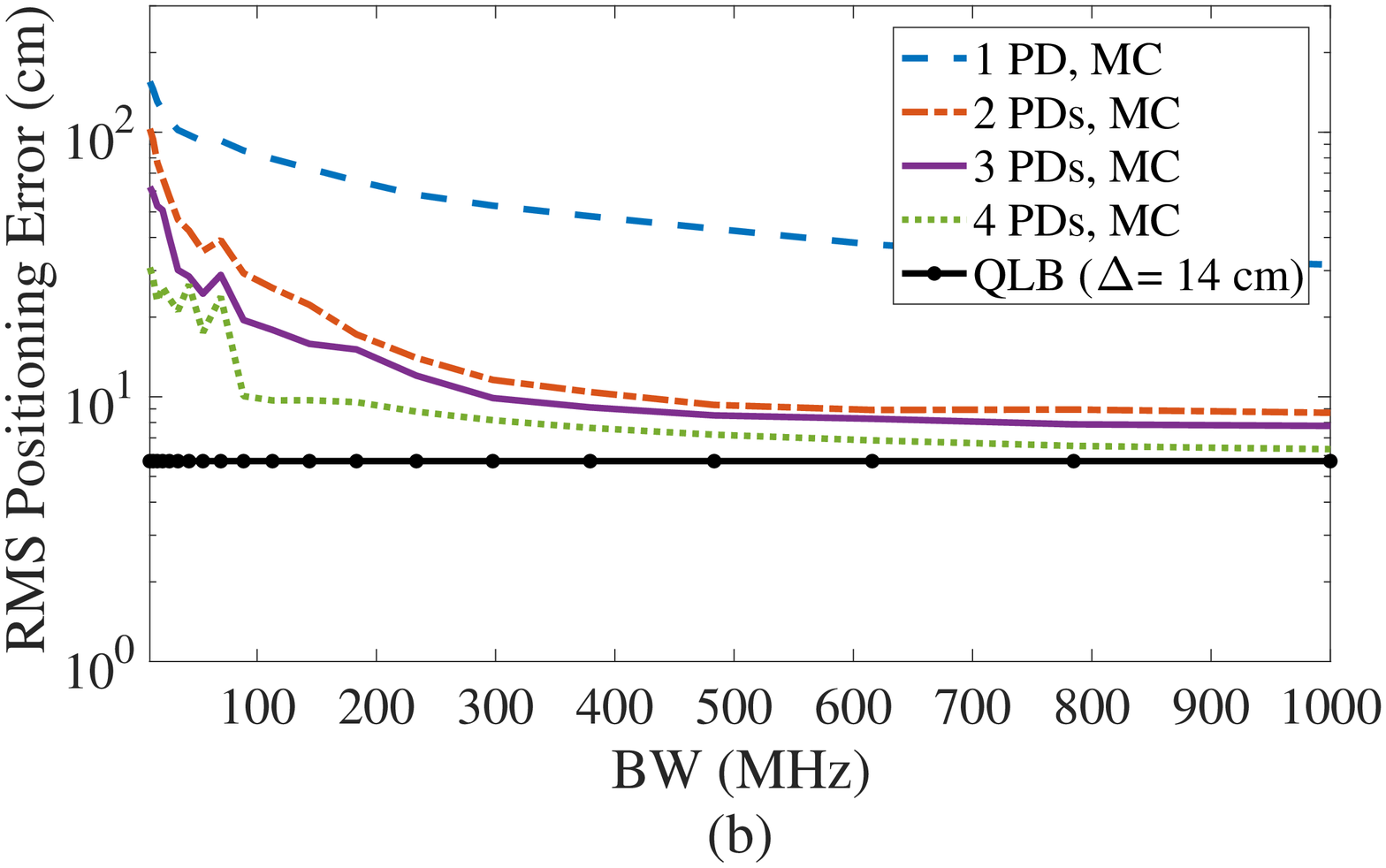} 
		\label{fig:BW_3obs}
	\end{subfigure}   
	\caption{RMS distance error vs. the transmitter BW, for grid step size of 14 cm and multiple PDs scenarios for a) two features algorithm, b) three features algorithm.}
	\label{fig:RMS_BW} 
\end{figure}

In practice, the system BW is limited by the LED BW. In this regard, we investigate the optical channel frequency response to see to what extent this lowpass filtering eliminates fingerprinting information and consequently degrades the localization performance. Fig. \ref{fig:FreqRespFDL} illustrates the channel frequency response amplitude corresponding to different locations of the user in the room. The DC bias in the frequency domain is due to the LOS received optical power strength. By removing the LOS and calculating the $3$ dB BW of the channel frequency response, we can see in Fig. \ref{fig:FreqRespFDL}  that the channel BW is less than 200 MHz for points far from the walls of the room. Fig. \ref{fig:BW_Surf} shows the optical channel $3$ dB BW for the whole room. The channel BW is less than 150 MHz in all regions except for the edges. Hence, a significant part of the fingerprint information of the optical channel, the SPP and the $\Delta \tau$, is included in the lower frequencies. Fig. \ref{fig:RMS_BW} illustrates the RMS positioning error for different system BWs. The results show promising accuracy for a typical IR LED bandwidth of 100 MHz. The accuracy is an almost monotonically increasing function of BW since the higher frequency resolution leads to more distinguishable constellation points in the observation space. The fluctuations in the low-frequency part of the plots are due to the fact that we miss some important parts of the channel information by LED low-pass filtering.


\section{Conclusion and Future Works}
\label{Concl}

In this paper, an uplink infrared positioning algorithm is introduced, exploiting the diffuse part of the optical channel for localization rather than considering it as a noise. Extracting the most informative components of the channel impulse response, a feature map of the room is created. This technique requires only one receiver-transmitter pair for localization, yet using additional PDs helps reduce the localization error. An expression of the CRLB for the proposed algorithm is presented that shows high potential for the proposed fingerprinting scheme. 
The numerical results show that the LED bandwidth limitation would not affect the algorithm, as the informative part of the optical channel characteristics is embedded in frequencies lower than $100$ MHz, which is in the range of expected BWs for off-the-shelf IR LEDs. 

The performance of the nearest neighbor estimation algorithm is evaluated for several different grid step sizes and multiple PDs scenarios. The results show an RMS positioning error of $5$~cm using 4 PDs, an SNR of $50$~\si{dB}, and a grid step size of $14$~\si{cm}, which is close to the quantization error lower bound. The theoretical analysis matches numerical results. 

As future work, this algorithm can be extended for practical scenarios that consider transmitter tilt, shadowing effects and more realistic channel models. Given that the fingerprints are room specific, learning the fingerprints map of the different locations is one of the practical challenges that can be addressed in future work.
Adding a motion tracking algorithm can further improve the performance of our localization algorithm \cite{7795368,rahaim2012state}.
\section{Acknowledgement}
This work was funded in part by the National Science Foundation (NSF) through the STTR program, under award number 1521387.

\section*{Appendix: Parametric Model of Channel Features}
In this section we explain the details of the regression-based closed-form approximations for two important parameters in the proposed positioning algorithm: the SPP, and the delay interval between LOS and SPP components.
\begin{figure*}
	\centering
	\begin{subfigure}{}
		\includegraphics[width=2in]{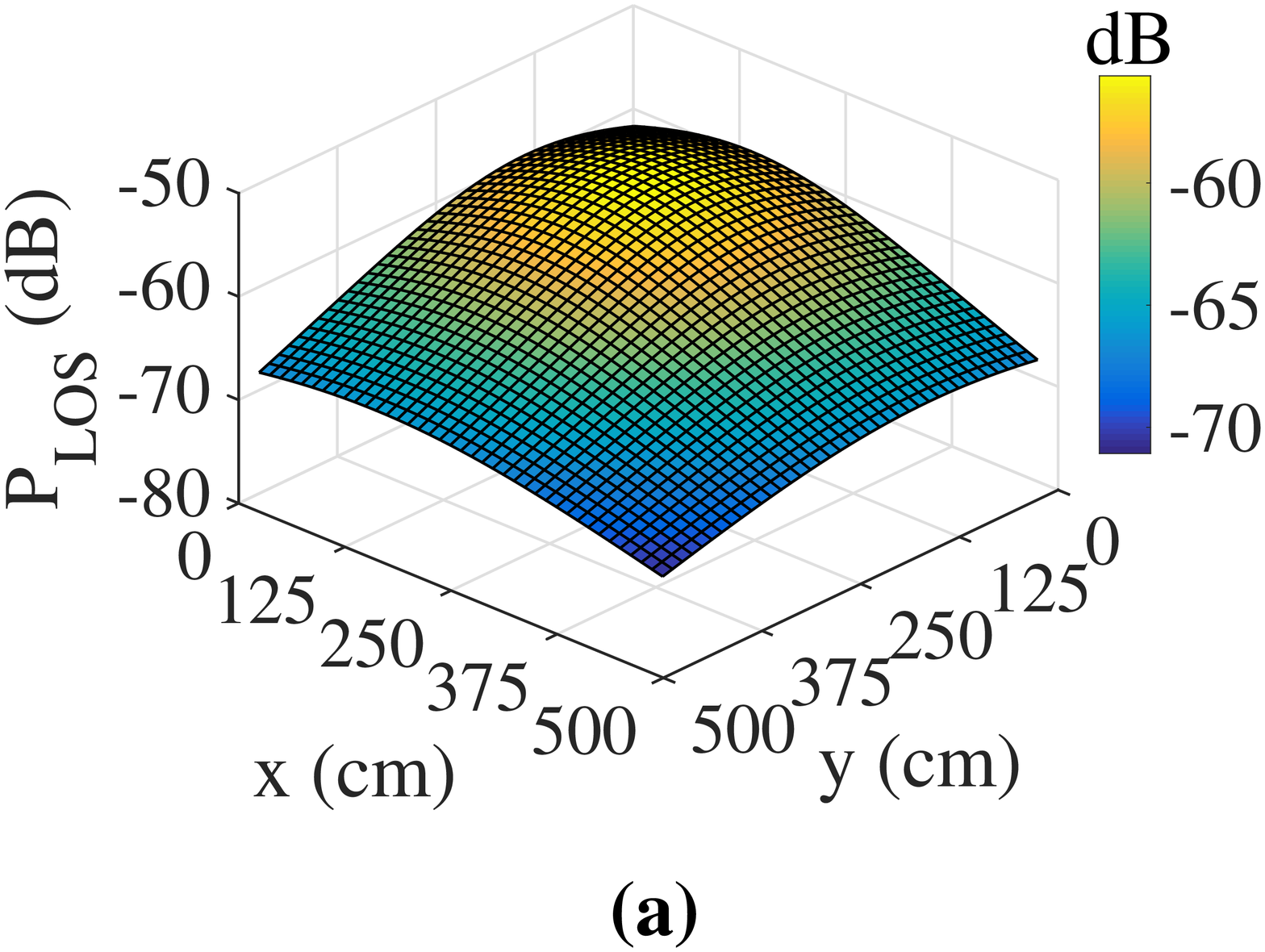}
		\label{fig:LOS}
	\end{subfigure}
	\begin{subfigure}{}
		\includegraphics[width=2in]{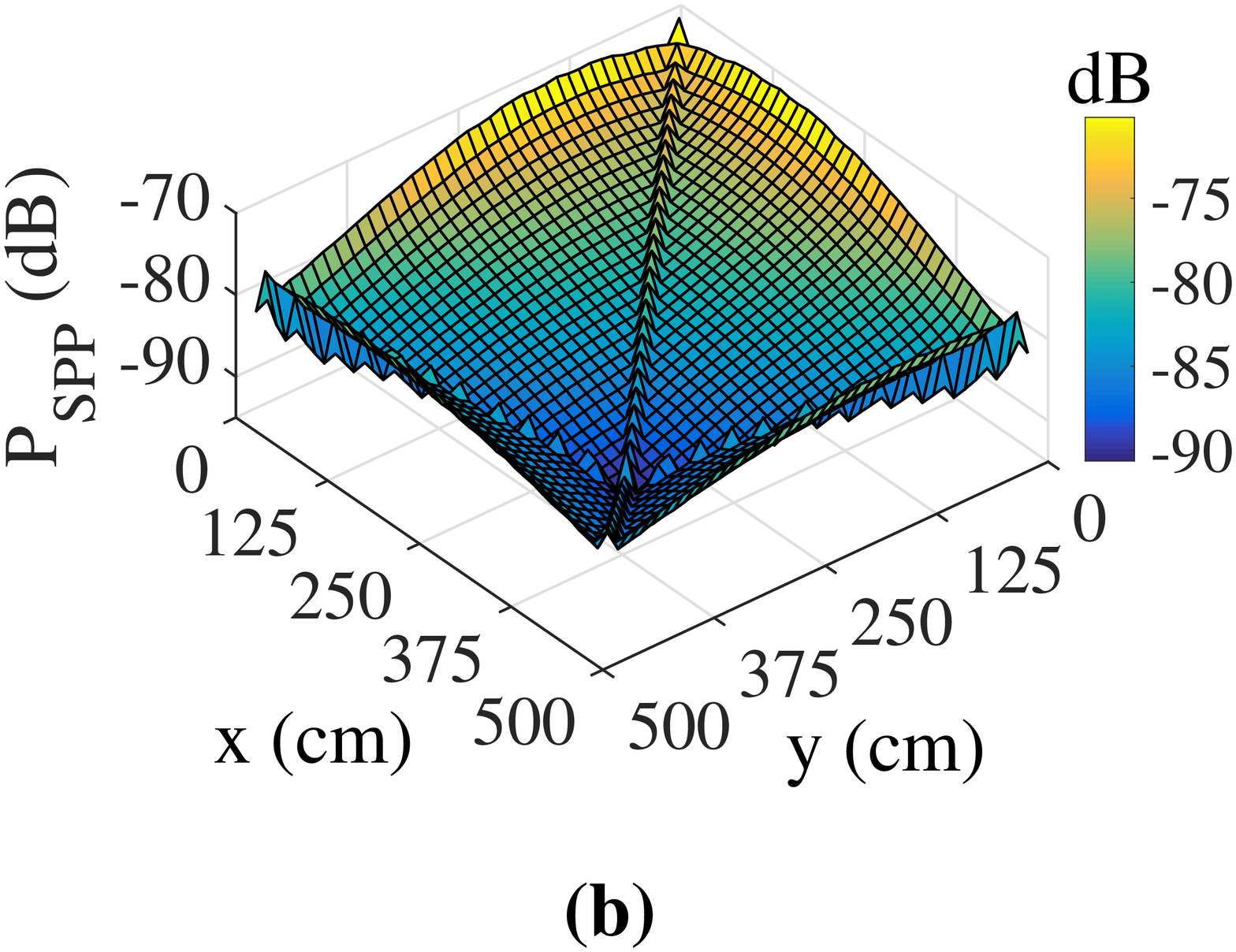} 
		\label{fig:SP}
	\end{subfigure}   
	\begin{subfigure}{}
		\includegraphics[width=2in]{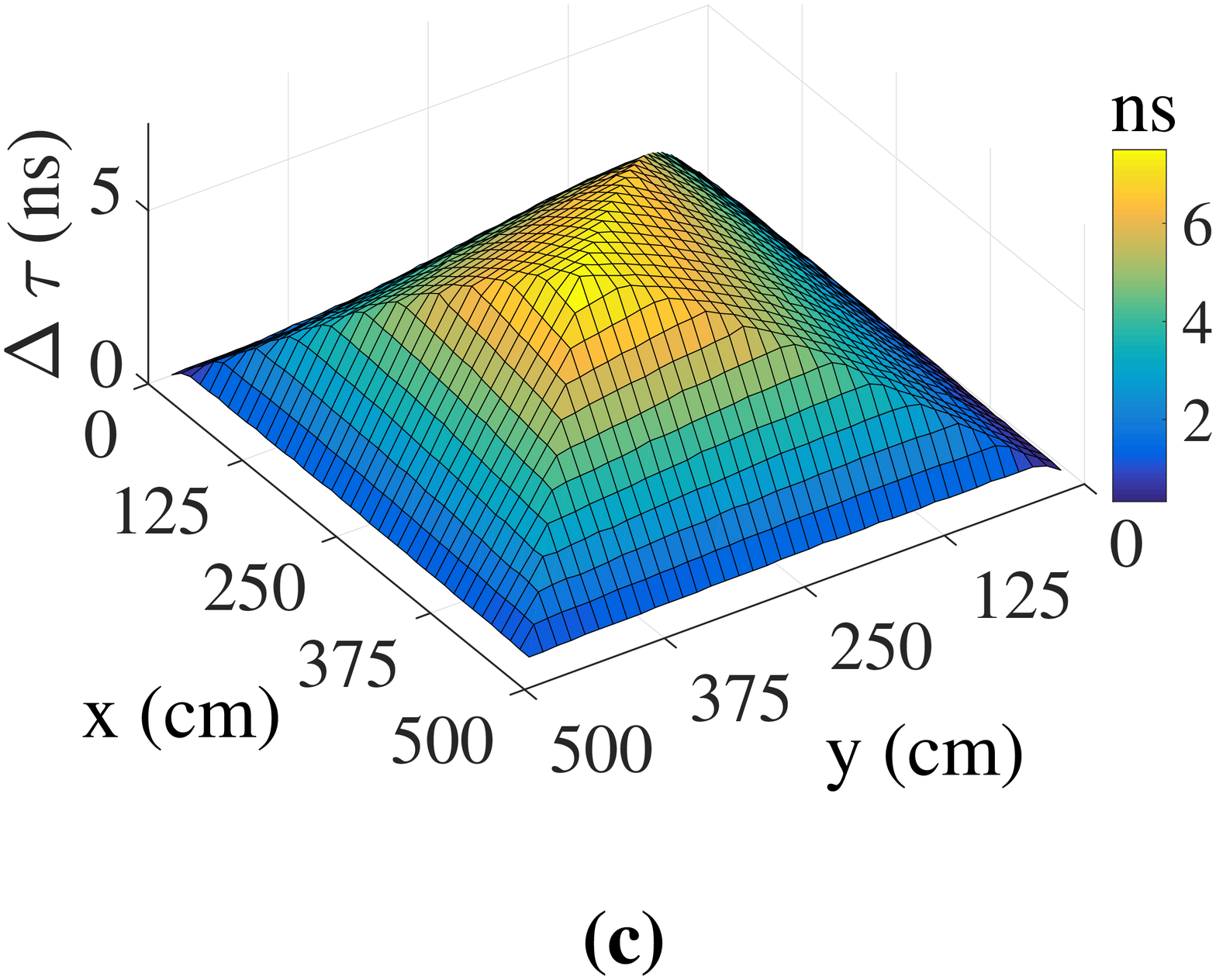} 
		\label{fig:SPtop}
	\end{subfigure}   
	\caption{a) LOS, and b) SPP components of received impulse response (dB), and c) Time delay between the LOS and SPP components (ns).}
	\label{fig:LOSSP} 
\end{figure*}
Fig. \ref{fig:LOSSP} illustrates the surface plot of extracted feature components of the channel impulse response for the room configuration described in Table~\ref{TableI} using PD1. 
\begin{figure}[h]
	\centering
	\includegraphics[width=2.8in]{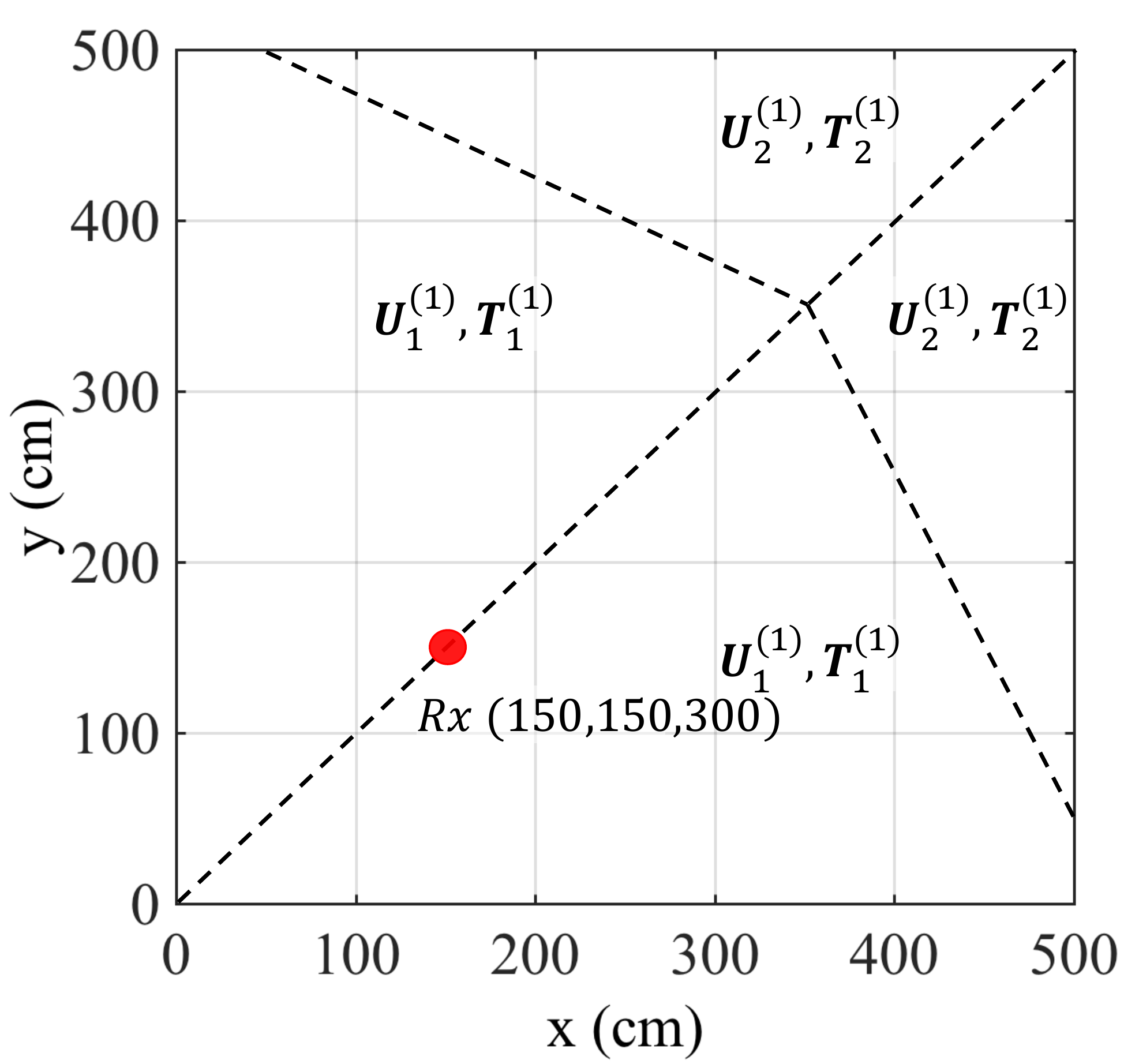} 
	\caption{Birds-eye view of the room.}
	\label{fig:Bird_eye}
\end{figure}

Considering the smoothness of the $P_{SPP}$ and $\Delta\tau$ surfaces, these can be approximated by using a polynomial regression. In order to reduce the order of the polynomial approximation, we divide each surface into 4 sections that can be approximated by using a polynomial surface of degree 4 in each variable using $\mathbf{A}(x)=\begin{bmatrix} 1 \ x \ x^2 \ x^3 \ x^4 \end{bmatrix}^T$, as in (\ref{CRB_apx11}). 
These sections are defined considering the symmetry and edges of the surfaces. In this particular example both of the SPP and the $\Delta\tau$ surfaces have the same edge, hence, the same sections. Fig. \ref{fig:Bird_eye} illustrates a birds-eye view of the room for the SPP and the $\Delta\tau$ surfaces. Using numerical simulation, $\mathbf{U}^{(1)}$ and $\mathbf{T}^{(1)}$ corresponding to the one receiver are obtained as
\begin{align*}
\mathbf{U}^{(1)}=\left\{\begin{matrix}
\mathbf{U}^{(1)}_1 ,& x<y, y>5-0.43x\\ 
\mathbf{U}^{(1)}_2 ,& x<y<5-0.43x
\end{matrix}\right.
\end{align*}
\begin{align*}
\mathbf{U}^{(1)}_1=10^{-9}\begin{bmatrix} 
60.6& 108 & -58 & 8.25 & -0.3 \\ 
-251 & -47.3 & 41 & -4.5 & 0 \\ 
254  & -41 & -0.8 & 0 & 0 \\ 
-71 & 8.8 & 0 & 0 & 0 \\ 
5.5  & 0 & 0 & 0 & 0
\end{bmatrix}
\end{align*}
\begin{align}
\mathbf{U}^{(1)}_2=10^{-9}\begin{bmatrix}
9.08e3   & 1.64e3 & 80 & -18.4 & 1.83\\ 
-1.03e4 & -1.3e3 & 10 & -2 & 0 \\ 
4.3e3  & 300 & -1.4 & 0 & 0 \\ 
-788 & -24.4 & 0 & 0 & 0 \\ 
53  & 0 & 0 & 0 & 0
\end{bmatrix} 
\tag{A.1}
\end{align}
\begin{align*}
\mathbf{T}^{(1)}=&\left\{\begin{matrix}
\mathbf{T}^{(1)}_1 ,& x<y, y>5-0.43x\\ 
\mathbf{T}^{(1)}_2 ,& x<y<5-0.43x
\end{matrix}\right.
\end{align*}
\begin{align*}
\mathbf{T}^{(1)}_1=10^{-11}&\begin{bmatrix}
-2.2& 6.5 & -4.7 & 1.43 & -0.14 \\ 
329 & 101 & 41 & 4.17 & 0 \\ 
-13.1  & 0.08 & 1.36 & 0 & 0 \\ 
19.9 & 1.24 & 0 & 0 & 0 \\ 
20.7  & 0 & 0 & 0 & 0
\end{bmatrix}
\end{align*}
\begin{align}
\mathbf{T}^{(1)}_2=10^{-9}&\begin{bmatrix}
-20.7   & 13.8 & -11.4 & 1.97 & -0.05\\ 
26.8 & 9.5 & 10 & -0.24 & 0 \\ 
-14.9  & -2.03 & 0.3 & 0 & 0 \\ 
2.78 & -0.01 & 0 & 0 & 0 \\ 
-0.15  & 0 & 0 & 0 & 0
\end{bmatrix}
\tag{A.2}
\end{align}

The argument in $\mathbf{A}(\cdot)$ is in meters and  $\mathbf{U}^{(1)}$ and $\mathbf{T}^{(1)}$ are in the appropriate units so that $P_{SPP}^{(1)}(\boldsymbol\theta)$ is in mW and $\Delta \tau^{(1)}(\boldsymbol\theta)$ is in ns. For the region where $x>y$, these matrices can be defined based on Fig. \ref{fig:Bird_eye}. The regressions approximating $P_{LOS}(\boldsymbol{\theta})$ and $\Delta \tau(\boldsymbol{\theta})$ for other receiver locations in Table \ref{TableI} are omitted for the sake of brevity.

\ifCLASSOPTIONcaptionsoff
\fi
\balance
\bibliographystyle{IEEEtran}
\bibliography{VLP}

\end{document}